\documentclass[aps,twocolumn,prb]{revtex4-2}
\usepackage{graphicx}
\usepackage{amsmath}
\usepackage{cases}

\draft 

\begin{document}


\title{Anderson localization of two-dimensional massless pseudospin-1 Dirac particles
in a correlated random one-dimensional scalar potential}

\author{Seulong Kim}
\affiliation{Department of Energy Systems Research and Department of Physics, Ajou University, Suwon 16499, Korea}
\author{Kihong Kim}
\email{khkim@ajou.ac.kr}
\affiliation{Department of Energy Systems Research and Department of Physics, Ajou University, Suwon 16499, Korea}
\affiliation{School of Physics, Korea Institute for Advanced Study, Seoul 02455, Korea}

\date{\today}

\begin{abstract}
We study theoretically Anderson localization of two-dimensional massless pseudospin-1 Dirac particles
in a random one-dimensional scalar potential. We focus explicitly on the effect of disorder correlations, considering
a short-range correlated dichotomous random potential at all strengths of disorder. We also consider a
$\delta$-function correlated random potential at weak disorder.
Using the invariant imbedding method, we calculate the localization length in a numerically precise way and analyze its dependencies
on incident angle, disorder correlation length, disorder strength, energy, wavelength and average potential
over a wide range of parameter values.
In addition, we derive analytical formulas for the localization length, which are very accurate in the weak and strong
disorder regimes.
From the Dirac equation, we obtain an expression for the effective wave impedance, using which
we explain several conditions for delocalization. We also deduce a condition under which the localization length vanishes.
For all cases considered, the localization length depends non-monotonically on the disorder correlation length
and diverges as $\theta^{-4}$ as the incident angle $\theta$ goes to zero.
As the disorder strength is varied from zero to infinity, we find that there appear three different scaling regimes.
As the energy or wavelength is varied from zero to infinity, there appear three or four different scaling regimes with
different exponents, depending on the value of the average potential.
The crossovers between different scaling regimes are explained in terms of the disorder correlation effect.
\end{abstract}

\maketitle

\section{introduction}

In Dirac materials, the quasiparticles obey an effective Dirac-type equation and their dynamics resembles that of relativistic particles.
Since the experimental isolation of single-layer graphene in 2004, the interest in Dirac materials has increased explosively
motivated by their promising potential applications in various devices and many interesting physical properties \cite{novo,weh}.
In addition to single-layer and bilayer graphene,
the study of Dirac materials has expanded to other pseudo-relativistic materials
such as topological insulators, pseudospin-$N$ Dirac materials, Weyl and Dirac semimetals
and Kane semiconductors and the list keeps growing \cite{kats,neto,rozh,rozh2,hasan,zawa,armi}.
Since Dirac materials and metamaterials can be realized in condensed-matter systems,
photonic-crystal structures and cold-atom optical lattices, researches in
pseudo-relativistic systems have become very popular in many disciplines of physics including
condensed matter physics, optics and atomic physics \cite{deng,guo,ozawa,leek,garr}.

If we ignore effects such as intervalley scattering, electron spin and electron-electron and electron-phonon interactions,
the quasiparticles in graphene are described by a two-dimensional (2D) pseudospin-1/2 Dirac equation for massless particles, where the pseudospin
represents the two different sublattices of the underlying honeycomb lattice \cite{kats}. More recently, the study of pseudospin-1/2 systems
has been extended to those with general pseudospin $N$ ($=1$, 3/2, 2, $\cdots$) \cite{dora,lan}.
In the massless case, all these pseudospin-$N$ Dirac systems display the Klein tunneling effect, which is
manifested as a total transmission
of normally-incident particles through an arbitrary scalar potential barrier \cite{kats2,been,nicol}.
In the case of pseudospin-1 systems, an omnidirectional total transmission phenomenon termed super-Klein tunneling,
which occurs when the particle energy
is precisely one half of the constant potential barrier, has attracted much interest \cite{shen,urban,fang0,bo1,kim_rip}. In addition to these,
there have been many recent theoretical studies exploring various properties of pseudospin-1 systems \cite{ley,wang}.

The unique transport properties of Dirac materials also influence the nature
of Anderson localization in random potentials \cite{nomura,liao,makh}.
Differences in the types of wave equations, the material properties and the nature of disorder
can influence Anderson localization strongly \cite{she,sch,mod,izra,gre,segev,bpn2,king}.
Since the wave equation obtained from the Dirac equation
is of a substantially different type from the Schr\"odinger equation and the electromagnetic wave equation,
we expect conceptually new localization phenomena to arise in Dirac materials.
Anderson localization in pseudospin-1/2 systems in a random one-dimensional (1D) scalar potential has been studied
theoretically by several authors \cite{zhu,bli,zhao,kkd,fang,fang2}.
It has been shown that localization is destroyed at normal incidence due to the Klein tunneling effect.
Close to the normal incidence, the localization length $\xi$ has been shown to scale as $\xi\propto \theta^{-2}$,
where $\theta$ is the incident angle.
As the disorder strength increases from zero to infinity, the localization length shows a non-monotonic behavior, such that it
initially decreases, attains a minimum value and then increases to infinity.
This type of counterintuitive delocalization effect induced by strong disorder has been interpreted in terms of effective impedance \cite{kkd}.

Anderson localization of pseudospin-1 Dirac particles in a random 1D scalar potential
was studied theoretically in two recent papers \cite{fang, fang2}.
In Ref.~\cite{fang}, the localization length for a random multilayer structure, where
all layers had the same thickness
and the potential in each layer was a random variable distributed uniformly in the range $[-W,W]$, was calculated
numerically using the transfer matrix method by averaging over 4000 random configurations. In addition,
some analytical results on the localization length were obtained using the surface Green function method.
The authors reported a very peculiar transition behavior such that
the localization length diverged
as $\xi\propto\sin^{-4}\theta$ at normal incidence when $W$ was smaller than the particle energy $E$, while
it diverged as $\xi\propto\sin^{-2}\theta$ when $W$ was larger than $E$.

In Ref.~\cite{fang2}, the localization length was calculated as a function of the wavelength in the long wavelength limit
using the same method as in Ref.~\cite{fang} for three different types of random multilayer models.
In the first model, the potential in each layer was randomly selected from the two values $V_0+W$ and $V_0-W$.
In this binary case, the authors showed that the localization length scaled as $\xi\propto \lambda^6$ if $V_0=0$
and as $\xi\propto \lambda^4$ if $V_0\ne 0$ in the long wavelength limit. When $V_0\ne 0$, they also reported that
there appeared a sharp peak of the localization length at the value of $\lambda$ corresponding to $E=V_0$,
signifying the onset of delocalization.
In the second model, the potential in the $i$-th layer was randomly selected from the two values $W(1+\delta_i)$ and $-W(1+\delta_i)$,
where $\delta_i$ is a random number distributed uniformly in a small interval $[-Q,Q]$.
In this case, the authors showed that $\xi$ scaled as $\xi\propto \lambda^4$ in the long wavelength limit.
In the third model, layers with zero potential were alternated periodically with those with a random potential and
the scaling of $\xi$ was found to be $\xi\propto \lambda^2$ in the long wavelength limit. It was argued that
this nonuniversal dependence of the scaling exponent on the specific type of the random potential was
a characteristic of pseudospin-1 Dirac systems.

We point out that the random multilayer structures considered
in Refs.~\cite{fang} and \cite{fang2} have short-range disorder correlations imbedded in them.
However, the influence of these correlations on localization phenomena was not studied explicitly.
In Ref.~\cite{fang}, the incident angle dependence was investigated carefully, but the scaling dependence on other parameters
was not analyzed in detail especially in the strong disorder regime.
On the other hand, the results reported in Ref.~\cite{fang2} were limited to the long wavelength limit.
Therefore it is highly desirable to study the scaling behavior of a single model
in the entire wavelength (or energy) range from zero to infinity
and the effect of disorder correlations on the the crossover between different scaling behaviors.

Recently, there has been much interest in the roles of short-range and long-range disorder correlations in localization
phenomena \cite{izra,21,22,23}. In this paper, we extend this approach to the localization in Dirac systems.
We consider the continuum Dirac equation in 2D for massless pseudospin-1 particles in a random 1D scalar potential
and calculate the localization length using the invariant imbedding method (IIM) \cite{kly,kim1,kim2,kim3,kim4,lee,kim6,kim5}.
In order to study the scaling behavior of the localization length in the entire energy range
and the influence of disorder correlations
on it, we consider a short-range correlated dichotomous (or binary) random potential which is explicitly characterized
by a correlation length $l_c$
as well as its strength. In addition, we consider a $\delta$-correlated (or uncorrelated) random potential in the weak disorder
regime and compare the results with those from the short-range correlated model. One of the main advantages of our IIM
is that we can perform the disorder averaging analytically in an exact way and convert the original random problem into
an equivalent nonrandom one. Therefore we can avoid repeating calculations for a large number of random configurations and then averaging
over the results. Another advantage is that it is possible to derive analytical formulas for the localization length which are
extremely accurate in the weak and strong disorder regimes.

Using our method, we investigate the dependencies of the localization length on incident angle, disorder correlation length, disorder strength,
energy, wavelength and average potential over a wide range of parameter values.
We find different scaling behaviors in the different regions of the parameter space
and explain the crossovers between them in terms of the disorder correlation effect. We also derive an expression
for the effective wave impedance, using which we interpret several delocalization phenomena.

Massless pseudospin-1 systems are characterized by two Dirac cones intersected by a flat band at the Dirac point. It has been known that
they can be realized in 2D lattices such as the Lieb, dice and kagome lattices \cite{flach}.
Recently, there has been much progress in the experimental realization of pseudospin-1 systems.
Several candidate 2D materials, which may be described by the pseudospin-1 Dirac equation,
have been discovered \cite{wli,lzhu}. There also have been many attempts to construct artificial 2D lattice structures displaying the properties
of pseudospin-1 systems, based on
cold-atom optical lattices, photonic crystal structures and artificial electronic lattices \cite{guz,mukh,slot}.
Due to this rapid development, we expect that it will be possible to explore the unique properties of these systems
experimentally in the near future.

The rest of this paper is organized as follows.
In Sec.~\ref{sec:model}, we introduce the pseudospin-1 Dirac equation and the two different kinds of random potentials used in this study.
We also derive an analytical expression for the wave impedance.
In Sec.~\ref{sec:iim}, the IIM for the calculation of the localization length is described and the invariant imbedding equations for the
two random models are derived. In Sec.~\ref{sec:analytic}, we apply the perturbation expansion method to the invariant imbedding equations
and derive analytical formulas for the localization length in the weak and strong disorder regimes.
In Sec.~\ref{sec:numerical}, we present detailed numerical results obtained using the IIM and discuss the dependencies of the localization length
on incident angle, disorder correlation length, disorder strength,
energy, wavelength and average potential. We conclude the paper in Sec.~\ref{sec:con}.

\section{Model}
\label{sec:model}

The effective Hamiltonian that describes massless pseudospin-1 Dirac-like particles moving in the 2D $xy$ plane
in a 1D scalar potential $U=U(x)$ takes the form
\begin{eqnarray}
{\mathcal H}=v_F \left(S_x p_x+S_y p_y\right)+UI,
\end{eqnarray}
where $v_F$ is the Fermi velocity and $I$ is the $3\times 3$ unity matrix.
The $x$ and $y$ components of the pseudospin-1 operator, $S_x$ and $S_y$, are represented by
\begin{eqnarray}
S_x=\frac{1}{\sqrt{2}}\begin{pmatrix} 0& 1& 0\\ 1& 0& 1\\ 0& 1& 0\end{pmatrix},~~
S_y=\frac{1}{\sqrt{2}}\begin{pmatrix} 0& -i& 0\\ i& 0& -i\\ 0& i& 0\end{pmatrix}.
\end{eqnarray}
In the case where the potential $U$ depends only on $x$,
the $x$ and $y$ components of the momentum operator, $p_x$ and $p_y$, are given by
\begin{eqnarray}
p_x=\frac{\hbar}{i}\frac{d}{dx},~~p_y=\hbar k_y,
\end{eqnarray}
where $k_y$ is the constant of motion.

The time-independent Dirac equation in 2D for the three-component vector wave function $\psi$ [$=\left( \psi_1, \psi_2, \psi_3\right)^{\rm T}$] is
\begin{eqnarray}
{\mathcal H}\psi=E\psi,
\end{eqnarray}
which is a set of three coupled first-order differential equations for $\psi_1$, $\psi_2$ and $\psi_3$.
We can eliminate $\psi_1$ and $\psi_3$ using the equations
\begin{eqnarray}
\psi_1&=&-\frac{i}{\sqrt{2}}\frac{\hbar v_F}{E-U}\left(\frac{d}{dx}+k_y\right)\psi_2,\nonumber\\
\psi_3&=&-\frac{i}{\sqrt{2}}\frac{\hbar v_F}{E-U}\left(\frac{d}{dx}-k_y\right)\psi_2,
\end{eqnarray}
and obtain a single wave equation for $\psi_2$ of the form
\begin{eqnarray}
&&\frac{d}{dx}\left(\frac{\hbar v_F}{E-U}\frac{d\psi_2}{dx}\right)+\left(\frac{E-U}{\hbar v_F}
-\frac{\hbar v_F{k_y}^2}{E-U}\right)\psi_2=0.
\label{eq:we1}
\end{eqnarray}

We assume that a plane wave described by $\psi_2$ is incident obliquely from the free region $x>L$ where $U=0$
onto the nonuniform region in $0\le x\le L$ where $U=U(x)$, and then transmitted to the free region $x<0$ where $U=0$.
The wave number in the free regions, $k$, is related to the particle energy $E$ by
\begin{eqnarray}
k=\frac{E}{\hbar v_F}.
\end{eqnarray}
The {\it negative} $x$ component of the wave vector in the incident and transmitted regions, $k_x$,
and the $y$ component of the wave vector, $k_y$, are given by
\begin{eqnarray}
k_x=k\cos\theta,~k_y=k\sin\theta,
\end{eqnarray}
where $\theta$ is the incident angle.

We introduce a dimensionless quantity $\epsilon=\epsilon(x)$ defined by
\begin{eqnarray}
\epsilon=1-u,
\end{eqnarray}
where $u=U/E$.
In the incident and transmitted regions where $U=0$, we have $\epsilon=1$.
In terms of $\epsilon$, the wave equation, Eq.~(\ref{eq:we1}), can be written concisely as
\begin{eqnarray}
\frac{d}{dx}\left(\frac{1}{\epsilon}\frac{d\psi_2}{dx}\right)+{k_x}^2\epsilon\eta^2\psi_2=0,
\label{eq:we0}
\end{eqnarray}
where $\eta$ is defined by
\begin{eqnarray}
\eta^2&=&\frac{1}{\cos^2\theta}\left(1-\frac{\sin^2\theta}{\epsilon^2}\right).
\label{eq:imped1}
\end{eqnarray}
We notice that the wave equation of this form looks identical to that for $p$-polarized electromagnetic
waves propagating normally in a
medium with the wave impedance function given by $\eta(x)$.
In the free regions where $\epsilon=1$, $\eta^2$ is identically equal to 1.
Therefore, if $\eta^2$ is unity in the inhomogeneous region $0\le x\le L$ as well, the impedance is matched throughout the space
and there will be no wave scattering.

There are two cases where the impedance matching can be achieved. If the incident angle $\theta$ is zero, $\eta^2$ is identically equal to 1
regardless of the magnitude and the functional form of $U(x)$. This is the well-known Klein tunneling phenomenon occurring at normal incidence.
The other case occurs when $\epsilon=-1$ corresponding to a constant potential barrier of height $U=2E$.
In this case, $\eta^2$ is unity for all $\theta$ and therefore an {\it omnidirectional} total transmission arises.
This phenomenon has been termed super-Klein tunneling \cite{shen,urban,fang0,bo1,kim_rip}.

In this paper, we are interested in the localization of pseudospin-1 Dirac-like particles in a random 1D scalar potential.
We assume that in the region $0\le x\le L$,
$U$ is a random function of $x$ given by
\begin{eqnarray}
U=U_0+\delta U(x),
\end{eqnarray}
where $U_0$ is the disorder-averaged value of $U$
and $\delta U(x)$ is the randomly-fluctuating part of $U$ with zero mean.
We will consider two different types of $\delta U(x)$.
In the first case, it is a Gaussian random function satisfying
\begin{eqnarray}
\langle\delta U(x)\delta U(x^\prime)\rangle=G\delta(x-x^\prime),
~~\langle\delta U(x)\rangle=0,
\end{eqnarray}
where the notation $\langle\cdots\rangle$ denotes averaging over disorder and $G$ is a
parameter characterizing the strength of disorder. This case corresponds to {\it uncorrelated} white noise.
In the second case, we assume that $\delta U(x)$ is a short-range correlated dichotomous (or binary) Gaussian
random function, which fluctuates randomly between
the two values $S$ and $-S$ and satisfies
\begin{equation}
\langle\delta U(x)\delta U(x^\prime)\rangle=S^2 \exp\left(-\vert x-x^\prime\vert/l_c\right),~~\langle\delta U(x)\rangle=0.
\end{equation}
In this case, $S^2$ measures the strength of disorder and the parameter $l_c$ is the correlation length of disorder.
By {\it short-range} correlation, we mean that the correlation function $\langle\delta U(x)\delta U(x^\prime)\rangle$
decays exponentially versus $\vert x-x^\prime\vert$. If it decays as a power law, the random potential is
called {\it long-range} correlated.

Our method is based on deriving a set of equivalent {\it nonrandom} differential equations
starting from the original random wave equation, where
the effect of randomness is taken care of analytically.
Later, we will show that in the case of $\delta$-correlated randomness, this method can be applied only
when the disorder is sufficiently weak, while in the case of dichotomous randomness, it can be applied to arbitrary strengths of disorder.
Therefore our numerical calculations will be mainly focused on the case of short-range correlated dichotomous randomness.

\section{Invariant imbedding method}
\label{sec:iim}

We can solve the wave equation, Eq.~(\ref{eq:we1}), using
the IIM. In this method, we first calculate the reflection and transmission coefficients $r$ and $t$ defined
by the wave functions in the incident and transmitted regions:
\begin{eqnarray}
\psi_2\left(x,L\right)=\left\{\begin{array}{ll}
  e^{-ik_x\left(x-L\right)}+r(L)e^{ik_x\left(x-L\right)}, & x>L \\
  t(L)e^{-ik_xx}, & x<0
  \end{array},\right.\nonumber\\
\end{eqnarray}
where $r$ and $t$ are regarded as functions of $L$.
The first step in the derivation of the invariant imbedding equations for $r$ and $t$ is to rewrite Eq.~(\ref{eq:we1}) in the form
 \begin{eqnarray}
\frac{d}{dx}\begin{pmatrix} f_1\\ f_2\end{pmatrix}=A\begin{pmatrix} f_1\\ f_2\end{pmatrix},
\end{eqnarray}
where $A$ ia a $2\times 2$ matrix.
With the definitions
\begin{eqnarray}
f_1=\psi_2,~~f_2=\frac{\hbar v_F}{E-U}\frac{d{\psi_2}}{dx},
\end{eqnarray}
we can easily obtain
\begin{eqnarray}
A=\begin{pmatrix} 0 & \frac{E-U}{\hbar v_F}\\ \frac{\hbar v_F{k_y}^2}{E-U}-\frac{E-U}{\hbar v_F} & 0\end{pmatrix}.
\end{eqnarray}
Then we follow the procedure described in Ref.~\cite{kim5} and derive
\begin{eqnarray}
&&\frac{1}{k}\frac{dr}{dl}=-\frac{i\cos\theta}{2}\epsilon\left(r-1\right)^2\nonumber\\&&~~~~~~~~~
+\frac{i}{2\cos\theta}\left(\epsilon-\frac{\sin^2\theta}{\epsilon}\right)\left(r+1\right)^2,\nonumber\\
&&\frac{1}{k}\frac{dt}{dl}=-\frac{i\cos\theta}{2}\epsilon\left(r-1\right)t\nonumber\\&&~~~~~~~~~
+\frac{i}{2\cos\theta}\left(\epsilon-\frac{\sin^2\theta}{\epsilon}\right)\left(r+1\right)t,
\label{eq:imbd1}
\end{eqnarray}
where $r$, $t$ and $\epsilon$ are functions of $l$.
Here, the variable $l$ represents the thickness of the system in the direction of the inhomogeneity and is called the imbedding parameter.
The quantity $r(l)$ denotes the reflection coefficient of a hypothetical system of thickness $l$ and $\epsilon(l)$ has the
same functional form as $\epsilon(x)$.

For any functional form of $U$ and for any values of $kL$ and $\theta$,
we can integrate these equations from $l=0$ to $l=L$ using the initial conditions $r(0)=0$ and $t(0)=1$
and obtain $r(L)$ and $t(L)$. The reflectance $R$ and the transmittance $T$ are obtained using $R=\vert r\vert^2$ and $T=\vert t\vert^2$.
In the absence of dissipation, the law of energy conservation requires that $R+T=1$.
In this paper, we are mainly interested in calculating the localization length $\xi$ defined by
\begin{equation}
\xi=-\lim_{L\to\infty}\left(\frac{L}{\langle\ln{T}\rangle}\right).
\label{eq:lld}
\end{equation}
The differential equation satisfied by $\ln T$ can be easily derived from the second of Eq.~(\ref{eq:imbd1}).

The parameter $\epsilon$ in Eq.~(\ref{eq:imbd1}) is a random function of $l$. Therefore these equations are
stochastic differential equations with random coefficients.
The nonrandom differential equations satisfied by the disorder-averaged quantities such as $\langle R\rangle$, $\langle T\rangle$
and $\langle \ln
T\rangle$ can be derived from Eq.~(\ref{eq:imbd1}) using standard methods of stochastic differential equations.
We will use Furutsu-Novikov formula \cite{fru,nov} in the case of $\delta$-correlated randomness and the formula of differentiation of
Shapiro and Loginov \cite{shap} in the
case of short-range correlated dichotomous randomness.

\subsection{$\delta$-correlated random potential}

In the case of a $\delta$-correlated random potential, we have difficulty in handling the random function $\epsilon$
appearing in the denominators of some coefficients in Eq.~(\ref{eq:imbd1}). It is possible to raise it to the numerator
if the randomness is sufficiently weak such that
\begin{eqnarray}
\bigg\vert \frac{\delta u}{\epsilon_0}\bigg\vert \ll 1,
\label{eq:approx}
\end{eqnarray}
where
\begin{eqnarray}
\epsilon_0=1-u_0,~~u_0=\frac{U_0}{E},~~\delta u=\frac{\delta U}{E}.
\end{eqnarray}
Then we can use the approximation
\begin{eqnarray}
\frac{1}{\epsilon}=\frac{1}{\epsilon_0-\delta u}\approx\frac{1}{\epsilon_0}\left(1+\frac{\delta u}{\epsilon_0}\right).
\end{eqnarray}

It is now straightforward to derive the equation for $\langle \ln T\rangle$ using Furutsu-Novikov formula, which takes the form
\begin{eqnarray}
&&-\frac{1}{k}\frac{d \langle\ln{T}\rangle}{dl}=\frac{g{D_1}^2}{\cos^2{\theta}}+\frac{\epsilon_0 C_1}{\cos\theta}{\rm Im}(Z_1)
\nonumber\\
&&~~~~~~~+\frac{2g D_0D_1}{\cos^2\theta}{\rm Re}(Z_1)+\frac{g {D_1}^2}{\cos^2\theta}{\rm Re}(Z_2),
\label{eq:ll1}
\end{eqnarray}
where $Z_n$ ($n=1,2,\cdots $) and the parameters $g$, $C_1$, $D_0$ and $D_1$ are defined by
\begin{eqnarray}
&&Z_n=\langle r^n\rangle,~~g=\frac{G}{4\hbar v_F E},~~
C_1=\left(1-\frac{1}{{\epsilon_0}^2}\right)\sin^2\theta,\nonumber\\
&&D_0=1+\cos^2\theta+\frac{\sin^2\theta}{{\epsilon_0}^2},~~D_1=\left(1+\frac{1}{{\epsilon_0}^2}\right)\sin^2\theta.\nonumber\\
\end{eqnarray}
From Eqs.~(\ref{eq:lld}) and (\ref{eq:ll1}), we find that the localization length $\xi$ is given by
\begin{eqnarray}
&&\frac{1}{k\xi}=\frac{g{D_1}^2}{\cos^2{\theta}}+\frac{\epsilon_0 C_1}{\cos\theta}{\rm Im}[Z_1(l\to\infty)]
\nonumber\\
&&~~~~~~+\frac{2g D_0D_1}{\cos^2\theta}{\rm Re}[Z_1(l\to\infty)]\nonumber\\
&&~~~~~~+\frac{g {D_1}^2}{\cos^2\theta}{\rm Re}[Z_2(l\to\infty)].
\label{eq:llz}
\end{eqnarray}

Using the first of Eq.~(\ref{eq:imbd1}) and Furutsu-Novikov formula, we can also derive an infinite number of coupled
differential equations for $Z_n$:
\begin{eqnarray}
&&\frac{1}{k}\frac{dZ_n}{dl}=\left[\frac{i\epsilon_0 }{\cos\theta}nC_0-\frac{g}{\cos^2\theta}n^2\left(2{D_0}^2+{D_1}^2\right)\right]Z_n\nonumber\\
&&~~~~~~+\left[\frac{i\epsilon_0 }{2\cos\theta}nC_1-\frac{g}{\cos^2\theta}n(2n+1)D_0D_1\right]Z_{n+1}\nonumber\\
&&~~~~~~+\left[\frac{i\epsilon_0 }{2\cos\theta}nC_1-\frac{g}{\cos^2\theta}n(2n-1)D_0D_1\right]Z_{n-1}\nonumber\\
&&~~~~~~-\frac{g}{2\cos^2\theta}n(n+1){D_1}^2Z_{n+2}\nonumber\\
&&~~~~~~-\frac{g}{2\cos^2\theta}n(n-1){D_1}^2Z_{n-2},
\label{eq:imz}
\end{eqnarray}
where $C_0$ is defined by
\begin{eqnarray}
C_0=1+\cos^2\theta-\frac{\sin^2\theta}{{\epsilon_0}^2}.
\end{eqnarray}
These equations are supplemented with the initial conditions $Z_0=1$ and $Z_n=0$ for $n\ge 1$.
In the large $l$ limit corresponding to the case where the thickness of the random region diverges, 
all $Z_n$'s should approach constants independent of $l$, if the disorder-averaged value of $\epsilon$ is a constant independent of $x$. 
Then we can set the left-hand side of Eq.~(\ref{eq:imz})
to zero and obtain an infinite number of coupled algebraic equations. We solve these equations numerically by a systematic truncation method \cite{kim1}
and obtain $Z_1(l\to\infty)$ and $Z_2(l\to\infty)$, using which we calculate the localization length.

\subsection{Short-range correlated dichotomous random potential}

In the case of a short-range correlated dichotomous random potential, the parameter $\epsilon$ takes only two values, $\epsilon_0+\sigma$ and
$\epsilon_0-\sigma$, where $\sigma$ is defined by $\sigma=S/E$.
Then $1/\epsilon$ fluctuates between $1/(\epsilon_0+\sigma)$ and $1/(\epsilon_0-\sigma)$, the average of which is 
$\epsilon_0/({\epsilon_0}^2-\sigma^2)$. From this, we obtain the identity
\begin{equation}
\frac{1}{\epsilon}=\frac{\epsilon_0}{{\epsilon_0}^2-\sigma^2}+\frac{\delta u}{{\epsilon_0}^2-\sigma^2}.
\end{equation}
In contrast to the $\delta$-correlated case, there is no approximation involved here.
The nonrandom differential equation satisfied by $\left\langle\ln{T}\right\rangle$ takes the form
\begin{equation}
-\frac{1}{k}\frac{d\langle\ln{T}\rangle}{dl}=\frac{1}{\cos{\theta}}{\rm Im}\left(\epsilon_0\tilde C_1Z_1-\tilde D_1W_1\right),
\label{eq:ll2}
\end{equation}
where $W_n$ ($n=1,2,\cdots$), $\tilde C_1$ and $\tilde D_1$ are defined by
\begin{eqnarray}
&&W_n=\langle r^n \delta u\rangle, \nonumber\\
&&\tilde C_1=\left(1-\frac{1}{{\epsilon_0}^2-\sigma^2}\right)\sin^2\theta,\nonumber\\
&&\tilde D_1=\left(1+\frac{1}{{\epsilon_0}^2-\sigma^2}\right)\sin^2\theta.
\label{eq:cof1}
\end{eqnarray}
The localization length is given by
\begin{equation}
\frac{1}{k\xi}=\frac{1}{\cos{\theta}}{\rm Im}{\left[\epsilon_0\tilde C_1Z_1(l\to\infty)-\tilde D_1W_1(l\to\infty)\right]}.
\end{equation}

In order to obtain $Z_1$ and $W_1$ in the large $l$ limit,
we need to derive an infinite number of coupled nonrandom differential equations satisfied by $Z_n$'s and $W_n$'s using the formula of differentiation of Shapiro and Loginov, which takes the form
\begin{eqnarray}
\frac{d\langle \zeta^j f \rangle}{dl} = \bigg\langle
\zeta^j\frac{df}{dl} \bigg\rangle - \frac{j}{l_c}\langle \zeta^j f
\rangle +\frac{j(j-1)}{l_c}\sigma^2\langle \zeta^{j-2}f \rangle,\nonumber\\
\label{eq:sl}
\end{eqnarray}
where $j$ is an arbitrary nonnegative integer and the function $f$
satisfies an ordinary differential equation with random
coefficients expressed in terms of $\zeta$ \cite{shap}.
The dichotomous Gaussian random function $\zeta$ satisfies the condition
\begin{equation}
\langle \zeta(l)\zeta(l^\prime)\rangle=\sigma^2 \exp\left(-\vert x-x^\prime\vert/l_c\right),~~\langle \zeta(l)\rangle=0.
\end{equation}
By substituting $f$ and $\zeta$ with $r^n$ and $\delta u$ respectively and taking $j=0,1$, we obtain
\begin{eqnarray}
&&\frac{\cos{\theta}}{ik}\frac{d Z_n}{dl}=n\epsilon_0\tilde C_0Z_n+\frac{1}{2}n\epsilon_0\tilde C_1\left(Z_{n+1}+Z_{n-1}\right)\nonumber\\
&&~~~~~~~~-n\tilde D_0W_n-\frac{1}{2}n \tilde D_1\left(W_{n+1}+W_{n-1}\right), \nonumber\\
&&\frac{\cos{\theta}}{ik}\frac{d W_n}{dl}=\left(n\epsilon_0\tilde C_0+\frac{i\cos{\theta}}{k l_c}\right)W_n\nonumber\\
&&~~~~~~~~+\frac{1}{2}n\epsilon_0\tilde C_1\left(W_{n+1}+W_{n-1}\right)\nonumber\\
&&~~~~~~~~-n\sigma^2 \tilde D_0Z_n-\frac{1}{2}n\sigma^2\tilde D_1\left(Z_{n+1}+Z_{n-1}\right),
\label{eq:imz2}
\end{eqnarray}
where
\begin{eqnarray}
&&\tilde C_0=1+\cos^2{\theta}-\frac{\sin^2{\theta}}{{\epsilon_0}^2-\sigma^2},\nonumber\\
&&\tilde D_0=1+\cos^2{\theta}+\frac{\sin^2{\theta}}{{\epsilon_0}^2-\sigma^2}.
\label{eq:cof2}
\end{eqnarray}
They are supplemented with the initial conditions $Z_0=1$, $Z_n=0$ for $n\ge 1$ and $W_n=0$ for all $n$.
In the large $l$ limit, all $Z_n$'s and $W_n$'s are constants independent of $l$. Then we can set the left-hand sides of Eq.~(\ref{eq:imz2})
to zero and obtain an infinite number of coupled algebraic equations. We solve these equations numerically by a systematic truncation method
and obtain $Z_1(l\to\infty)$ and $W_1(l\to\infty)$, using which we calculate the localization length.

\section{Analytical expressions for the localization length in the weak and strong disorder regimes}
\label{sec:analytic}

Though we can solve the invariant imbedding equations, Eqs.~(\ref{eq:ll1}), (\ref{eq:imz}), (\ref{eq:ll2}) and (\ref{eq:imz2}),
numerically for general parameter values, it is highly instructive to apply the perturbation theory to them to derive
analytical expressions for the localization length in some limiting cases.

\subsection{Weak disorder regime in a $\delta$-correlated random potential}
\label{sec:dw}

In the case of a $\delta$-correlated Gaussian random potential, the imbedding equations, Eqs.~(\ref{eq:ll1}) and (\ref{eq:imz}), have been
derived assuming that the disorder is sufficiently weak. Therefore we apply the perturbation theory to those equations
only in the weak disorder regime expressed by Eq.~(\ref{eq:approx}). We write the reflection coefficient in the large $l$ limit, $r$, as $r=r_0+\delta r$,
where $r_0$ is the reflection coefficient from an interface between free space and a half-space nonrandom medium
with the parameter $\epsilon_0$. The expression for $r_0$ takes the form
\begin{eqnarray}
r_0= \frac{\epsilon_0\cos\theta-\tilde p}{\epsilon_0\cos\theta+\tilde p},
\label{eq:r0}
\end{eqnarray}
where $\tilde p$ is defined by
\begin{eqnarray}
\tilde p=\left\{\begin{matrix} \mbox{sgn}(\epsilon_0)\sqrt{{\epsilon_0}^2-\sin^2\theta} &\mbox{if }{\epsilon_0}^2\ge \sin^2\theta \\
i\sqrt{\sin^2\theta-{\epsilon_0}^2} & \mbox{if }{\epsilon_0}^2< \sin^2\theta \end{matrix}.\right.
\end{eqnarray}

We expand $Z_n$ as
\begin{eqnarray}
Z_n=\langle\left(r_0+\delta r\right)^n\rangle=\sum_{j=0}^n {n\choose j} {r_0}^{n-j}\langle\left(\delta r\right)^j\rangle.
\label{eq:zex}
\end{eqnarray}
Using this and the infinite number of algebraic equations obtained from Eq.~(\ref{eq:imz}),
we get an infinite number of coupled equations for $\langle \left(\delta r\right)^n\rangle$ for all integers $n$.
We expand these averages in terms of the small perturbation parameter $g$.
From analytical considerations and numerical calculations, we can show that
the leading terms for $\langle \delta r\rangle$ and $\langle \left(\delta r\right)
^2 \rangle$ are of the first order in $g$, whereas
that for $\langle \left(\delta r\right)^3 \rangle$ is of the second order in $g$,
except at incident angles close to the critical angle of total
reflection $\theta_c$ satisfying ${\epsilon_0}^2= \sin^2\theta_c$. Therefore, we substitute
\begin{eqnarray}
&&Z_1=r_0+\langle \delta r\rangle,~Z_2={r_0}^2+2r_0\langle \delta r\rangle+\langle \left(\delta r\right)
^2 \rangle,\nonumber\\
&&Z_3\approx {r_0}^3+3{r_0}^2\langle \delta r\rangle+3r_0\langle \left(\delta r\right)^2 \rangle
\end{eqnarray}
into Eq.~(\ref{eq:imz}) when $n=1,2$  in the large $l$ limit
and obtain two coupled equations for $\langle \delta r\rangle$ and $\langle \left(\delta r\right)^2 \rangle$.
We solve these equations analytically and substitute the resulting expressions into Eq.~(\ref{eq:llz})
to the leading order in $g$. From this, we obtain the localization length of the form
\begin{eqnarray}
\frac{1}{k\xi}&=&2\sqrt{\sin^2\theta-{\epsilon_0}^2}~\Theta\left(\sin^2\theta-{\epsilon_0}^2\right)\nonumber\\&&
+\frac{4g \sin^4\theta}{{\epsilon_0}^2\left({\epsilon_0}^2-\sin^2\theta\right)},
\label{eq:af1}
\end{eqnarray}
where $\Theta$ is the step function, $\Theta(x)=1$ for $x>0$ and 0 for $x<0$.
We remind that the weak disorder expansion cannot be applied if $\epsilon_0\approx 0$ (that is, $E\approx U_0$)
or ${\epsilon_0}^2\approx \sin^2\theta$.

From Eq.~(\ref{eq:af1}), it follows that
there is a symmetry under the sign change of $\epsilon_0$ and $\theta$.
We also notice that in the total reflection (or tunneling)
regime where $\vert\sin\theta\vert>\vert\epsilon_0\vert$ and when the disorder parameter $g$ is sufficiently small, $\xi$
increases as $g$ increases from zero.
This is an example of the well-known disorder-enhanced tunneling phenomenon \cite{kkd,frei,luck,kim_t,hein,kim7}.
On the other hand, if $\vert\epsilon_0\vert$ is
greater than $\vert\sin\theta\vert$, we have $\xi\propto g^{-1}$, which simply means that localization is enhanced by weak disorder.

When $\theta$ is very close to zero, we find that $k\xi\approx [{\epsilon_0}^4/(4g)]\theta^{-4}$, which diverges at $\theta=0$
due to the Klein tunneling effect.
This $\theta^{-4}$ dependence of the localization length near $\theta=0$ is a unique characteristic of pseudospin-1 systems
and is distinct from the $\theta^{-2}$ dependence occurring in pseudospin-1/2 systems.
If the average potential $U_0$ is zero, we obtain
\begin{equation}
k\xi=\frac{1}{4g\sin^2\theta \tan^2\theta}.
\end{equation}
This result can be compared with
the dependence $\xi\propto\sin^{-4}\theta$ reported in Ref.~\cite{fang}, which has been derived using the transfer matrix method
and the surface Green function method for a random multilayer model,
when the average potential is zero and the disorder strength
is smaller than a certain critical value. In that model, all layers have the same thickness
and the potential in each layer is a random variable distributed uniformly in the range $[-W,W]$.
In Ref.~\cite{fang}, it has also been reported that $\xi\propto\sin^{-2}\theta$ if
the disorder strength $W$ is larger than the critical value given by $W_c=E$. Our method for the $\delta$-correlated random potential
cannot be applied to the strong disorder regime. However, we can solve exactly the case of a short-range correlated
dichotomous random potential for arbitrary strengths of disorder. In Secs.~\ref{sec:dw2} and \ref{sec:ds}, we will show that
the $\theta^{-4}$ dependence near $\theta=0$ is valid in that model regardless of the strength of disorder
and the size of the correlation length.

Next, we consider the dependence of the localization length on the energy of the incident particle, or
equivalently, on the frequency and the wavelength of the incident wave.
For that purpose, it is more convenient to normalize $\xi$ by the wave number associated with the random potential, $k_u$,
defined by
\begin{eqnarray}
k_u=\frac{G}{\left(\hbar v_F\right)^2}=4kg,
\label{eq:disk0}
\end{eqnarray}
which is independent of energy.
For simplicity, let us consider only the case where the average potential $U_0$ is zero.
Then it is trivial to show that
\begin{equation}
k_u\xi=\frac{1}{\sin^2\theta \tan^2\theta},
\label{eq:ed1}
\end{equation}
which is independent of energy (and, equivalently, of frequency and wavelength).
This result is applied in the weak disorder limit, which corresponds to the large energy (or frequency) or short wavelength limit.
However, we will show in the next subsection that the above result is an artifact of the $\delta$-correlated random model,
the spectral density of which has no ultraviolet cutoff, and more realistic short-range correlated random models
exhibit different behavior in the limit of asymptotically large energy.

\subsection{Weak disorder regime in a short-range correlated dichotomous random potential}
\label{sec:dw2}

In the case of a short-range correlated dichotomous random potential,
the wave number in the medium with dichotomous randomness where $\epsilon=\epsilon_0\pm\sigma$ is given by
$\epsilon_0\left(1\pm\sigma/\epsilon_0\right) k$. Therefore it is natural to represent the strength of disorder by the parameter
$s$ defined by
\begin{equation}
s=\frac{\sigma^2}{{\epsilon_0}^2},
\end{equation}
where $\epsilon_0\ne 0$.
In the weak disorder regime where $s$ is sufficiently small,
we can rewrite $\tilde C_0$, $\tilde C_1$ , $\tilde D_0$ and $\tilde D_1$ in Eqs.~(\ref{eq:cof1}) and (\ref{eq:cof2}) as
\begin{eqnarray}
\tilde C_0&\approx& 1+\cos^2{\theta}-\frac{\sin^2{\theta}}{{\epsilon_0}^2}\left(1+s\right),\nonumber\\
\tilde C_1&\approx& \sin^2{\theta}-\frac{\sin^2{\theta}}{{\epsilon_0}^2}\left(1+s\right),\nonumber\\
\tilde D_0&\approx& 1+\cos^2{\theta}+\frac{\sin^2{\theta}}{{\epsilon_0}^2}\left(1+s\right),\nonumber\\
\tilde D_1&\approx& \sin^2{\theta}+\frac{\sin^2{\theta}}{{\epsilon_0}^2}\left(1+s\right).
\end{eqnarray}

Similarly to Sec.~\ref{sec:dw}, we write $r=r_0+\delta r$, where $r_0$ is given by Eq.~(\ref{eq:r0}), and expand $Z_n$ as in Eq.~(\ref{eq:zex})
and $W_n$ as
\begin{eqnarray}
W_n=\langle\left(r_0+\delta r\right)^n\delta u\rangle=\sum_{j=1}^n {n\choose j} {r_0}^{n-j}\langle \left(\delta r\right)^j\delta u\rangle.
\end{eqnarray}
From analytical considerations and numerical calculations, we can show that both $\vert\langle(\delta r)^j\rangle\vert$
and $\vert\langle (\delta r)^j\delta u\rangle\vert$ decrease rapidly as $j$ increases. We can also demonstrate that the leading terms for $\langle\delta r\rangle$, $\langle(\delta r)^2\rangle$ and $\langle \delta r\delta u\rangle$ are of the first order in $s$, while those for $\langle(\delta r)^3\rangle$ and $\langle (\delta r)^2\delta u\rangle$ are of the second order in $s$.
From these considerations, we derive the analytical expression for the localization length in the weak disorder regime of the form
\begin{widetext}
\begin{subnumcases}
  {\frac{1}{k\xi}=}
         2\sqrt{\sin^2{\theta}-{\epsilon_0}^2}
         +\frac{\sigma^2\left(\sqrt{\sin^2{\theta}-{\epsilon_0}^2}-2{\epsilon_0}^2kl_c\right)\sin^2{\theta}}
         {{\epsilon_0}^2\left(\sin^2{\theta}-{\epsilon_0}^2\right)\left(1+2kl_c\sqrt{\sin^2{\theta}-{\epsilon_0}^2}\right)},& $\mbox{when } \sin^2{\theta}>{\epsilon_0}^2$, \label{eq:bwka}\\
         \frac{2kl_c\sigma^2\sin^4{\theta}}{{\epsilon_0}^2\left({\epsilon_0}^2
         -\sin^2{\theta}\right)\left[1+4k^2{l_c}^2\left({\epsilon_0}^2-\sin^2{\theta}\right)\right]}, &$\mbox{when } \sin^2{\theta}<{\epsilon_0}^2$.
      \label{eq:bwkb}
  \end{subnumcases}
\end{widetext}
From this, we find that there is a symmetry under the sign change of $\epsilon_0$ and $\theta$ and
the localization length diverges as $\theta^{-4}$ at $\theta=0$.
In the total reflection
regime where $\vert\sin\theta\vert>\vert\epsilon_0\vert$, the localization length
increases as the disorder strength $\sigma$ increases, only if
\begin{eqnarray}
\sqrt{\sin^2\theta-{\epsilon_0}^2}<2{\epsilon_0}^2kl_c.
\label{eq:ie1}
\end{eqnarray}
Therefore, in contrast to the case of $\delta$-correlated randomness,
the disorder-enhanced tunneling phenomenon in the present case occurs only when the correlation length
is sufficiently large and $\vert\epsilon_0\vert$ is not too small.

We also observe that when $\vert\epsilon_0\vert$ is larger than $\vert\sin\theta\vert$, the expression given by Eq.~(\ref{eq:bwkb})
reduces to Eq.~(\ref{eq:af1}) in the limit where $kl_c\to 0$ and $kl_c\sigma^2\to 2g$.
Therefore, except in the total reflection regime,
the case of $\delta$-correlated randomness can be considered as that of short-range correlated randomness
in the limit where the correlation length vanishes.

When $\vert\epsilon_0\vert$ is larger than $\vert\sin\theta\vert$, we have the dependence $\xi\propto \sigma^{-2}$ regardless of
the values of other parameters including the correlation length.
In addition, we have a non-monotonic dependence of $\xi$ on the correlation length $l_c$.
When $l_c$ is sufficiently small, $\xi$ is inversely proportional to $l_c$, whereas when
$l_c$ is sufficiently large, it is proportional to $l_c$, as we can see clearly from the approximate expressions derived from Eq.~(\ref{eq:bwkb}):
\begin{subnumcases}
{k\xi=}
\frac{{\epsilon_0}^2\left({\epsilon_0}^2-\sin^2{\theta}\right)}{2kl_c\sigma^2\sin^4{\theta}}, ~~~~~~~~ \mbox{if } l_c\ll l_c^{\rm min}, \label{eq:llap1a}\\
\frac{2{\epsilon_0}^2\left({\epsilon_0}^2-\sin^2{\theta}\right)^2kl_c}{\sigma^2\sin^4{\theta}},~ \mbox{if } l_c\gg l_c^{\rm min}, \label{eq:llap1b}
\end{subnumcases}
where $l_c^{\rm min}$ is the value of $l_c$ at which
the localization length takes a minimum value and is given by
\begin{eqnarray}
kl_c^{\rm min}=\frac{1}{2\sqrt{{\epsilon_0}^2-\sin^2\theta}}.
\label{eq:llap2}
\end{eqnarray}
When the average potential is zero, these expressions are simplified to
\begin{subnumcases}
{k\xi=}
\frac{1}{2kl_c\sigma^2\sin^2{\theta}\tan^2{\theta}}, &$ \mbox{if } l_c\ll l_c^{\rm min}$, \\
\frac{2kl_c}{\sigma^2\tan^4{\theta}}, & $\mbox{if } l_c\gg l_c^{\rm min}$.
\end{subnumcases}

Since $k\sqrt{{\epsilon_0}^2-\sin^2\theta}$ is the $x$ component of the wave vector in the random region in an averaged sense,
we can rewrite Eq.~(\ref{eq:llap2}) as
\begin{eqnarray}
\frac{l_c^{\rm min}}{\lambda_{\rm eff}}=\frac{1}{4\pi},
\label{eq:llap3}
\end{eqnarray}
where we have defined an effective wavelength $\lambda_{\rm eff}$ by
\begin{eqnarray}
\lambda_{\rm eff}=\frac{2\pi}{k\sqrt{{\epsilon_0}^2-\sin^2\theta}}.
\end{eqnarray}
Therefore the two different regimes described by Eqs.~(\ref{eq:llap1a}) and (\ref{eq:llap1b})
are distinguished by the relative size of the correlation length and the effective wavelength.
In the regime where the effective wavelength is much larger than the correlation length, the model is effectively
uncorrelated and the correlation effect can be ignored.
In the opposite regime where the effective wavelength is much smaller than the correlation length,
the correlation effect should be important.

We next consider the dependence of the localization length on the energy and the wavelength.
In the present model, we normalize $\xi$ by the wave number associated with the random potential, $k_d$,
defined by
\begin{eqnarray}
k_d=\frac{S}{\hbar v_F}=k\sigma,
\label{eq:disk01}
\end{eqnarray}
which is independent of energy.
For simplicity, we consider only the case where the average potential is zero such that $\epsilon_0=1$.
Then we obtain
\begin{equation}
k_d\xi=\frac{1+4\left(k/k_d\right)^2{k_d}^2{l_c}^2\cos^2\theta}{2k_dl_c\sin^2\theta \tan^2\theta},
\end{equation}
which can be approximated in two forms
\begin{widetext}
\begin{subnumcases}
{k_d\xi=}
\frac{1}{2k_dl_c\sin^{2}{\theta}\tan^{2}{\theta}}, &$ \mbox{if } 1\ll\frac{k}{k_d}\ll \frac{1}{2k_dl_c\cos{\theta}}$, \label{eq:ed2a}\\
\frac{2k_dl_c}{\tan^{4}{\theta}}\left(\frac{k}{k_d}\right)^2,  &$\mbox{if } \frac{k}{k_d}\gg \frac{1}{2k_dl_c\cos{\theta}}$, \label{eq:ed2b}
\end{subnumcases}
\end{widetext}
depending on the relative size of $k/k_d$ ($= E/S$).
This result is applied in the weak disorder limit, corresponding to the large energy (or frequency) or short wavelength limit.
We notice that Eq.~(\ref{eq:ed2a}), which is independent of energy, is the same as the corresponding
expression in the $\delta$-correlated case, Eq.~(\ref{eq:ed1}), since the parameter $k_u$ is equal to $2{k_d}^2l_c$
if we identify $kl_c\sigma^2$ with $2g$.
In the asymptotically large energy limit, however, we have to apply Eq.~(\ref{eq:ed2b}), which shows
that $\xi$ has the dependence $\xi\propto k^2$, $E^2$, $\omega^2$, $\lambda^{-2}$, where $\omega$ is the frequency and $\lambda$
is the wavelength. It is straightforward to verify that similar scaling behaviors are obtained when the average potential is nonzero.

The boundary between the two scaling regions is given by the condition $kl_c\cos\theta=1/2$.
We find that in the region described by Eq.~(\ref{eq:ed2a}), the disorder correlation effect is unimportant, while in the asymptotically large
energy region described by Eq.~(\ref{eq:ed2b}), correlations play a crucial role.
In a previous paper on the localization of pseudospin-1/2 Dirac electrons in a $\delta$-correlated random scalar potential, we
have reported that the localization length is independent of energy in the large energy limit. This result was due to the use of a
$\delta$-correlated random potential and the true behavior in the asymptotically large energy limit for more realistic
short-range correlated random models has to be the same $\xi\propto E^2$ behavior as obtained here.
If the correlation length is very small, however, a wide regime where $\xi$ is independent of energy and wavelength
should precede the asymptotic regime.

\subsection{Strong disorder regime in a short-range correlated dichotomous random potential}
\label{sec:ds}

In the strong disorder regime where $s\gg 1$, we can rewrite $\tilde C_0$, $\tilde C_1$ , $\tilde D_0$ and $\tilde D_1$ in Eqs.~(\ref{eq:cof1}) and (\ref{eq:cof2}) as
\begin{eqnarray}
\tilde C_0&\approx& 1+\cos^2{\theta}+\frac{\sin^2{\theta}}{{\epsilon_0}^2}\left(\frac{1}{s}+\frac{1}{s^2}\right),\nonumber\\
\tilde C_1&\approx& \sin^2{\theta}+\frac{\sin^2{\theta}}{{\epsilon_0}^2}\left(\frac{1}{s}+\frac{1}{s^2}\right),\nonumber\\
\tilde D_0&\approx& 1+\cos^2{\theta}-\frac{\sin^2{\theta}}{{\epsilon_0}^2}\left(\frac{1}{s}+\frac{1}{s^2}\right),\nonumber\\
\tilde D_1&\approx& \sin^2{\theta}-\frac{\sin^2{\theta}}{{\epsilon_0}^2}\left(\frac{1}{s}+\frac{1}{s^2}\right).
\end{eqnarray}
We note that in this perturbation regime, the parameter $\epsilon_0\approx 0$ is allowed.
We write $r=r_\infty+\delta r$, where $r_\infty$
is the reflection coefficient from an interface between free space and an
infinitely-disordered half-space medium
with the parameter $\epsilon_0$ and
is given by
\begin{eqnarray}
r_{\infty}=\frac{\cos{\theta}-1}{\cos{\theta}+1}.
\end{eqnarray}

Similarly to Sec.~\ref{sec:dw2}, we expand $Z_n$ and $W_n$ as
\begin{eqnarray}
&&Z_n=\langle\left(r_\infty+\delta r\right)^n\rangle=\sum_{j=0}^n {n\choose j} {r_\infty}^{n-j}\langle\left(\delta r\right)^j\rangle, \nonumber\\
&&W_n=\langle\left(r_\infty+\delta r\right)^n\delta u\rangle=\sum_{j=1}^n {n\choose j} {r_\infty}^{n-j}\langle \left(\delta r\right)^j\delta u\rangle.\nonumber\\
\end{eqnarray}
We can show that both $\vert\langle(\delta r)^j\rangle\vert$
and $\vert\langle (\delta r)^j\delta u\rangle\vert$ decrease rapidly as $j$ increases. We can also demonstrate
the scaling relationships
\begin{eqnarray}
&&\langle\delta r\rangle,~ \langle\delta r\delta u\rangle \propto s^{-1},\nonumber\\
&&\langle\left(\delta r\right)^2\rangle,~ \langle\left(\delta r\right)^2\delta u\rangle \propto s^{-2},\nonumber\\
&&\langle\left(\delta r\right)^3\rangle,~ \langle\left(\delta r\right)^3\delta u\rangle \propto s^{-3},\nonumber\\
&&\langle\left(\delta r\right)^4\rangle,~ \langle\left(\delta r\right)^4\delta u\rangle \propto s^{-4}.
\end{eqnarray}
From these considerations, we can derive an analytical expression of the localization length in the strong disorder regime of the form
\begin{eqnarray}
k\xi=\frac{\sigma^2}{2kl_c\sin^4{\theta}}\left(1+\frac{4k^2{l_c}^2\sigma^4}{{\epsilon_0}^2}\right).
\label{eq:bwk2}
\end{eqnarray}

Again, there is a symmetry under the sign change of $\epsilon_0$ and $\theta$.
The localization length depends on $\theta$ as $\xi\propto \sin^{-4}\theta$ for all parameter values.
It also diverges as $\epsilon_0\to 0$, $l_c\to\infty$ or $\sigma\to\infty$.
The divergence of $\xi$ in the strong disorder limit, which also occurs in the pseudospin-1/2 case, is counterintuitive and
can be understood from the form of the wave impedance, Eq.~(\ref{eq:imped1}).
We find that in the strong disorder limit where $\delta u$ is statistically much larger than 1, the impedance approaches to a constant given by $\eta\approx \vert\cos\theta\vert^{-1}$. Therefore, as the disorder strength approaches to infinity,
the system becomes effectively nonrandom and the localization length should diverge for all $\theta$.

We can also understand the delocalization occurring when $\epsilon_0$ vanishes from the form of the wave impedance.
In that case, $\epsilon^2$ is always equal to $\sigma^2$
and nonrandom. Therefore the impedance is nonrandom for any incident angle and complete delocalization arises.
However, we stress that this type of delocalization is {\it strictly limited to the dichotomous randomness} and will not arise
in more generic models.

It is easy to understand the delocalization occurring when $l_c\to\infty$.
In a short-range correlated model, the potential can be considered nonrandom within the correlation length $l_c$.
Therefore in the limit where $l_c$ diverges, the potential is effectively nonrandom and localization is destroyed.

Similarly to the weak disorder case, we have a non-monotonic dependence of $\xi$ on the correlation length $l_c$,
as we can see from the approximate expressions derived from Eq.~(\ref{eq:bwk2}):
\begin{subnumcases}
{k\xi=}
\frac{\sigma^2}{2kl_c\sin^4{\theta}}, & $\mbox{if } l_c\ll \tilde l_c^{\rm min}$, \label{eq:llap3a}\\
\frac{2kl_c\sigma^6}{{\epsilon_0}^2\sin^4{\theta}}, & $\mbox{if } l_c\gg \tilde l_c^{\rm min}$, \label{eq:llap3b}
\end{subnumcases}
where $\tilde l_c^{\rm min}$ is the value of $l_c$ at which
the localization length takes a minimum value and is
given by
\begin{eqnarray}
k\tilde l_c^{\rm min}=\frac{\vert\epsilon_0\vert}{2\sigma^2}.
\label{eq:llap4}
\end{eqnarray}
This condition can also be interpreted in terms of the effective wavelength.
If we take the absolute value of the geometric average of the wave numbers in the dichotomous random potential, $(\epsilon_0+\sigma)k$ and
$(\epsilon_0-\sigma)k$, as the effective wave number $k_{\rm eff}$, we obtain
$k_{\rm eff}=k\vert\sigma^2-{\epsilon_0}^2\vert /{\vert\epsilon_0\vert}$. From this,
we can define the effective wavelength in the strong disorder regime as $\tilde\lambda_{\rm eff}=2\pi\vert\epsilon_0\vert/(k\sigma^2)$.
Then Eq.~(\ref{eq:llap4}) can be written similarly to Eq.~(\ref{eq:llap3}).

We notice that in the regime where $l_c$ is sufficiently small, $\xi$ is proportional to $\sigma^2$, while
in the regime where $l_c$ is sufficiently large, $\xi$ is proportional to $\sigma^6$.
The $\xi\propto \sigma^6$ scaling behavior and the crossover between the two scaling regimes have never been
obtained before. We stress that these contrasting scaling behaviors are universal and
occur regardless of the value of the average potential $U_0$, unless $U_0=E$.
We also point out that $\tilde l_c^{\rm min}$ is inversely proportional to $\sigma^2$,
and therefore in the strong disorder regime, the minimum of the localization length occurs at a rapidly-decreasing value of $kl_c$
as $\sigma$ increases.

We next consider the dependence of the localization length on the energy and the wavelength.
We normalize $\xi$ by the wave number associated $k_d$
defined by Eq.~(\ref{eq:disk01}) and rewrite Eq.~(\ref{eq:bwk2}) as
\begin{widetext}
\begin{eqnarray}
k_d\xi=\frac{1}{2k_dl_c\sin^4\theta}\left(\frac{k}{k_d}\right)^{-4}\left[1+\frac{4{k_d}^2{l_c}^2}
{\left(\frac{k}{k_d}-\tilde u_0\right)^2}\right],
\label{eq:disk02}
\end{eqnarray}
\end{widetext}
where $\tilde u_0$ is the normalized average potential defined by $\tilde u_0=U_0/S$.
In the strong disorder regime, it is necessary to distinguish the cases where the average potential is zero or nonzero.
We first consider the case where $\tilde u_0=0$.
Then $k_d\xi$ can be approximated in two forms
\begin{widetext}
\begin{subnumcases}
{k_d\xi=}
\frac{2k_dl_c}{\sin^{4}{\theta}}\left(\frac{k}{k_d}\right)^{-6},  &$\mbox{if } \frac{k}{k_d}\ll 2k_dl_c$, \label{eq:ed3a}\\
\frac{1}{2k_dl_c\sin^{4}{\theta}}\left(\frac{k}{k_d}\right)^{-4}, &$ \mbox{if } 2k_dl_c\ll\frac{k}{k_d}\ll 1$, \label{eq:ed3b}
\end{subnumcases}
\end{widetext}
depending on the relative size of $k/k_d$.
This result is applied in the strong disorder limit, corresponding to the small energy (or frequency) or long wavelength limit.
The boundary between the two scaling regions is given by the condition $k=2{k_d}^2l_c$, which is equivalent to $\tilde\lambda_{\rm eff}=4\pi l_c$.
We find that in the region described by Eq.~(\ref{eq:ed3b}), the disorder correlation effect is unimportant, while in the asymptotically small
energy region described by Eq.~(\ref{eq:ed3a}), correlations play a crucial role and
 $\xi$ has the dependence $\xi\propto k^{-6}$, $E^{-6}$, $\omega^{-6}$, $\lambda^{6}$.
If the correlation length is extremely small, then a wide regime given by Eq.~(\ref{eq:ed3b}), where $\xi\propto k^{-4}$, $E^{-4}$, $\omega^{-4}$, $\lambda^{4}$, should precede the asymptotic regime.

On the other hand, if the average potential $\tilde u_0$ is nonzero, we can approximate Eq.~(\ref{eq:disk02}) as
  \begin{equation}
  k_d\xi=\frac{1+4\left(k_dl_c/\tilde u_0\right)^2}{2k_dl_c\sin^{4}{\theta}}\left(\frac{k}{k_d}\right)^{-4},
  \end{equation}
which applies if $k/k_d\ll \vert \tilde u_0\vert$.
Therefore, in this case, we have the dependence $\xi\propto k^{-4}$, $E^{-4}$, $\omega^{-4}$, $\lambda^{4}$ in the asymptotically small energy
or long wavelength limit.
In Ref.~\cite{fang2}, from a numerical study of a binary random multilayer model, the $\xi\propto\lambda^6$ ($\xi\propto\lambda^4$)
scaling behavior was obtained in the long wavelength limit,
when the average potential was zero (nonzero). These results agree with ours in the asymptotic long wavelength limit. However, the crossover between the two behaviors, which would occur when the average potential was zero, was not investigated in Ref.~\cite{fang2}.
An important advantage of our result is that different scaling behaviors for different values of the average potential
are incorporated in a single formula, Eq.~(\ref{eq:disk02}).

\section{Numerical results}
\label{sec:numerical}

\subsection{Incident angle dependence}

\begin{figure}
\centering\includegraphics[width=8.5cm]{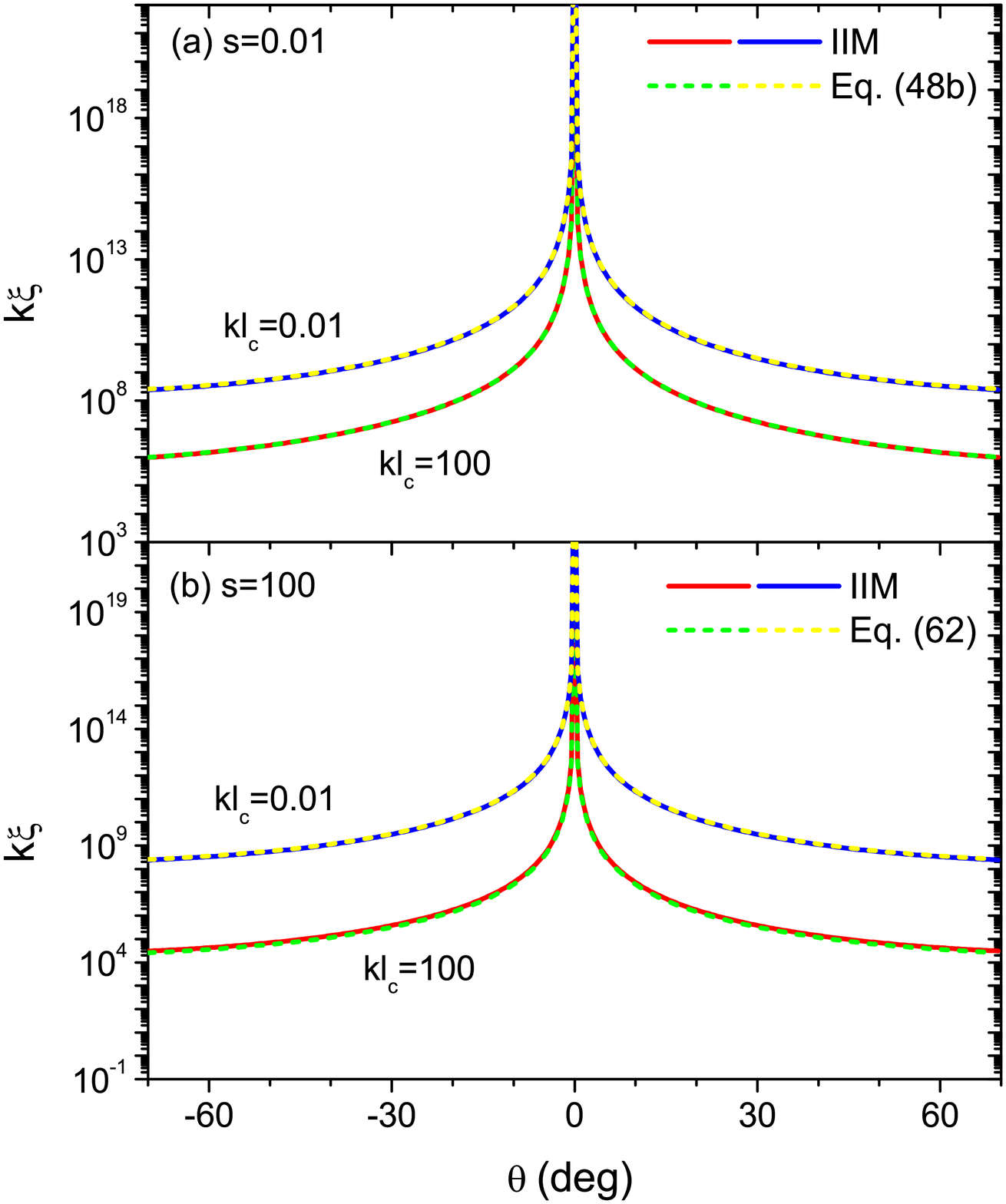}
 \caption{Normalized localization length $k\xi$ plotted versus incident angle $\theta$ in (a) the weak disorder regime with
 $s=0.01$ and (b) the strong disorder regime with $s=100$, when the average potential $U_0$ is zero and
 the normalized correlation length $kl_c$ is equal to 0.01 and 100.
 The numerical results obtained using the IIM are compared with the analytical formulas in the weak and strong disorder regimes, Eqs.~(\ref{eq:bwkb})
 and (\ref{eq:bwk2}).}
 \label{fig:1}
 \end{figure}

\begin{figure}
\centering\includegraphics[width=8.5cm]{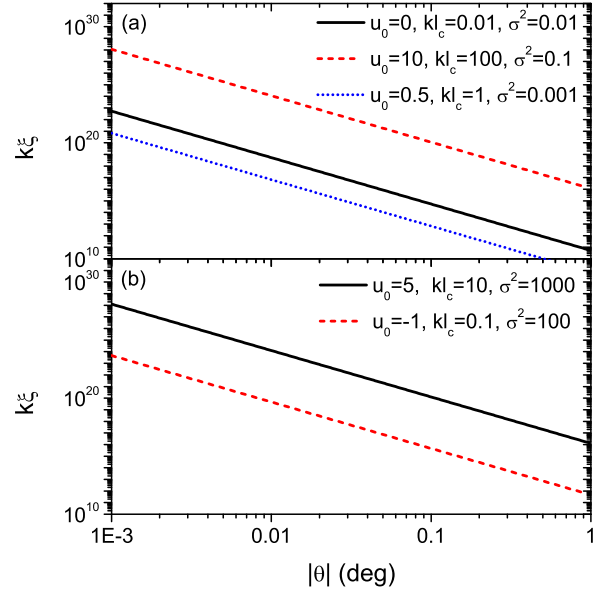}
 \caption{Normalized localization length $k\xi$ plotted versus $\vert\theta\vert$ in a log-log plot
 in the (a) weak and (b) strong disorder regimes. The five curves are obtained for different values of the parameters $u_0$, $kl_c$ and $\sigma^2$, which are indicated on the figure. All curves show the divergent behavior $k\xi\propto\theta^{-4}$ near $\theta=0$.}
 \label{fig:2}
\end{figure}

\begin{figure}
\centering\includegraphics[width=8.5cm]{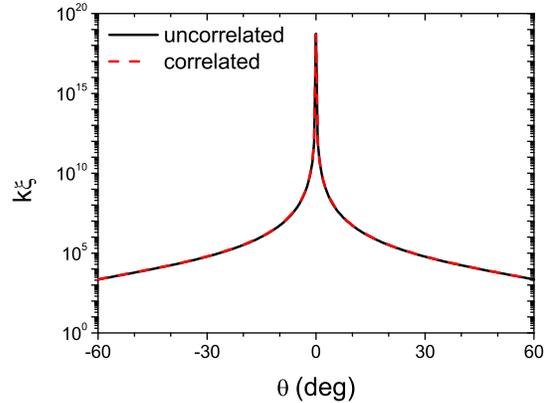}
\caption{Normalized localization length $k\xi$ plotted versus incident angle $\theta$ in the weak disorder regime when $U_0=0$. The solid curve corresponds to the $\delta$-correlated (or uncorrelated) case with $g=0.00005$, while the dashed one corresponds to the short-range correlated case with $kl_c=0.01$ and $\sigma^2=0.01$.}
\label{fig:3}
\end{figure}

In this section, we present the results of our comprehensive numerical calculations obtained using the IIM and discuss
the dependencies of the localization length on various parameters such as incident angle, disorder correlation length, disorder strength,
energy and wavelength. We first consider the incident angle dependance.

In Fig.~\ref{fig:1}, we consider a short-range correlated dichotomous random potential
and plot the normalized localization length $k\xi$ as a function of the incident angle $\theta$
in the weak disorder regime with $s=0.01$ and the strong disorder regime with $s=100$, when the average value of the potential $U_0$ is zero and
the normalized correlation length $kl_c$ is equal to 0.01 and 100.
The numerical results obtained using the IIM are compared with the analytical formulas in the weak and strong disorder regimes,
Eqs.~(\ref{eq:bwkb}) and (\ref{eq:bwk2}). The agreements are perfect.
All curves show that the localization length diverges rapidly as $\theta$ approaches to zero and complete delocalization
occurs at normal incidence.

In Fig.~\ref{fig:2}, we plot the normalized localization length $k\xi$ versus $\vert\theta\vert$ in a log-log plot
in the weak and strong disorder regimes. The curves shown were obtained for different values of the parameters $u_0$, $kl_c$ and $\sigma^2$.
We find that all curves show the same divergent scaling behavior $k\xi\propto\theta^{-4}$ near $\theta=0$ regardless of the parameter values.

We have also performed a numerical calculation for a $\delta$-correlated random potential in the weak disorder regime
and found the similar $\theta^{-4}$ scaling behavior.
In Fig.~\ref{fig:3}, we plot $k\xi$ versus $\theta$ when the average potential $U_0$ is zero and the disorder parameter
in the $\delta$-correlated case $g$ is equal to
0.00005 and compare it with the result obtained for the short-range correlated case with $kl_c=0.01$ and $\sigma^2=0.01$.
For the parameters chosen to satisfy the condition $kl_c\sigma^2=2g$, we find that the agreement between the two results is perfect.

\subsection{Disorder correlation length dependence}

\begin{figure}
\centering\includegraphics[width=8.5cm]{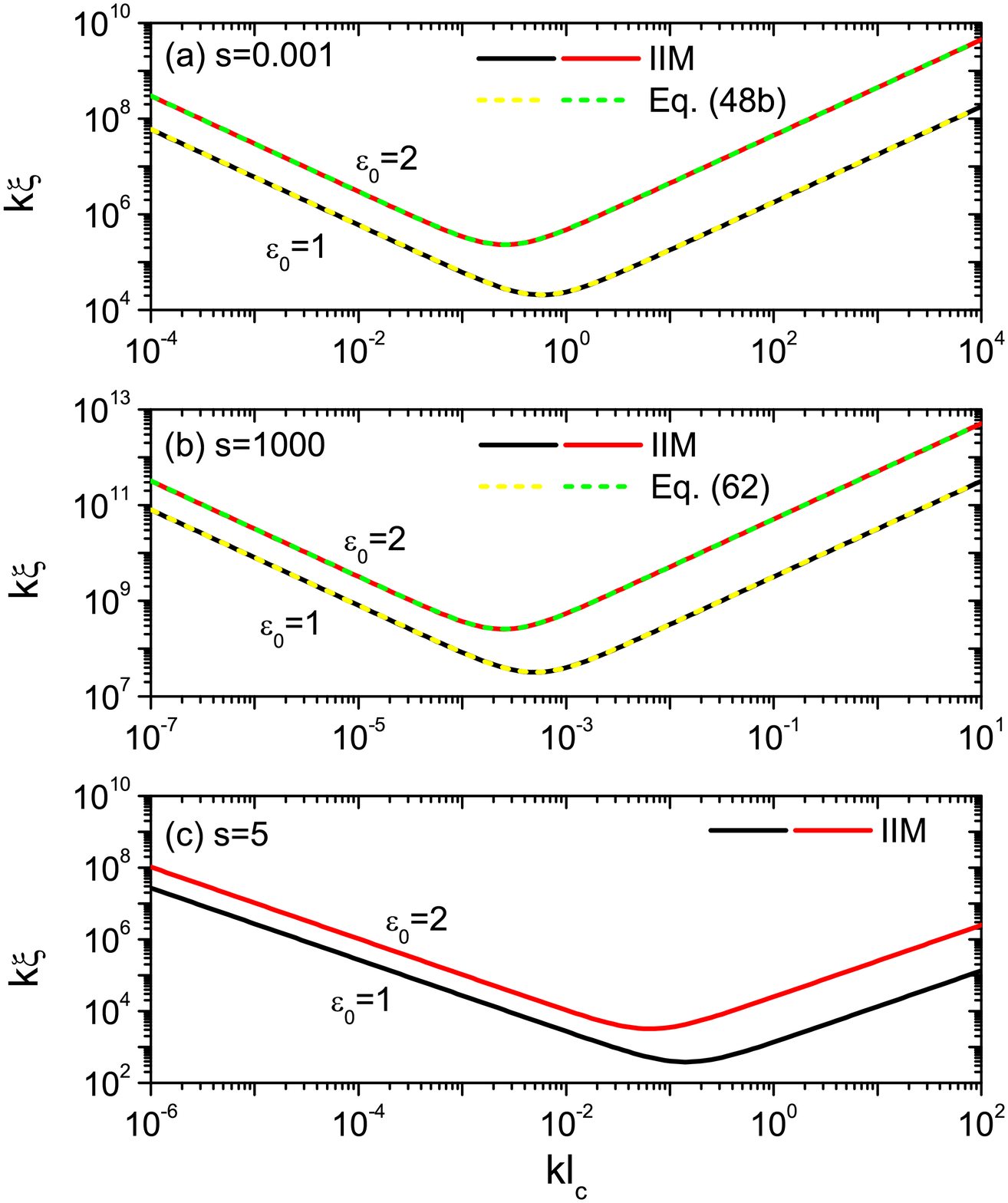}
 \caption{Normalized localization length $k\xi$ plotted versus normalized correlation length $kl_c$ in a log-log plot,
 when $\epsilon_0=1,2$, $\theta=30^\circ$ and (a) $s=0.001$, (b) $s=1000$, (c) $s=5$. In (a) and (b),
 the numerical results obtained using the IIM are compared with the analytical formulas, Eqs.~(\ref{eq:bwkb})
 and (\ref{eq:bwk2}), respectively.}
\label{fig:4}
\end{figure}

It is well-known that disorder correlations play a significant role in localization phenomena.
In Fig.~\ref{fig:4}, we plot the normalized localization length $k\xi$ versus normalized correlation length $kl_c$ in a log-log plot,
when the incident angle $\theta$ is fixed to $30^\circ$ and $\epsilon_0$ ($=1-u_0$) is 1 or 2.
We consider the weak disorder regime where $s=0.001$,
the strong disorder regime with $s=1000$ and the intermediate disorder regime where $s=5$. In all cases, the localization
length shows a non-monotonic dependence on $l_c$. As $l_c$ increases from zero to infinity,
the localization length initially decreases as $\xi\propto {l_c}^{-1}$, attains
a minimum value at $l_c^{\rm min}$ or $\tilde l_c^{\rm min}$ given by Eqs.~(\ref{eq:llap2})
and (\ref{eq:llap4}), and then increases as $\xi\propto {l_c}$.
In the weak and strong disorder regimes,
the numerical results obtained using the IIM agree very well with the analytical formulas, Eqs.~(\ref{eq:bwkb})
and (\ref{eq:bwk2}), respectively.
We have checked that in the weak disorder regime, the value of $l_c$ at which $\xi$ takes a minimum, $l_c^{\rm min}$,
is independent of the disorder strength $s$ (or $\sigma$).
As $s$ increases, however, this value decreases gradually and, in the strong disorder regime, $\tilde l_c^{\rm min}$ behaves as
$\tilde l_c^{\rm min}\propto \sigma^{-2}$ as given in Eq.~(\ref{eq:llap4}).

\subsection{Disorder strength dependence}

\begin{figure}
\centering\includegraphics[width=8.5cm]{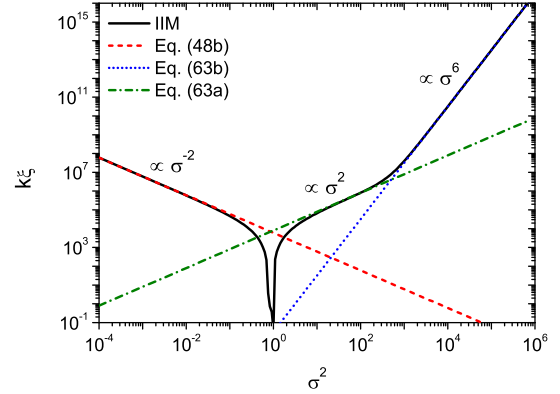}
 \caption{Normalized localization length $k\xi$ plotted versus disorder strength $\sigma^2$ in a log-log plot, when $kl_c=0.001$, $U_0=0$ and $\theta=30^\circ$. The numerical results obtained using the IIM are compared with the analytical formulas, Eqs.~(\ref{eq:bwkb}),
 (\ref{eq:llap3a}) and (\ref{eq:llap3b}), respectively.}
 \label{fig:5}
\end{figure}

\begin{figure}
\centering\includegraphics[width=8.5cm]{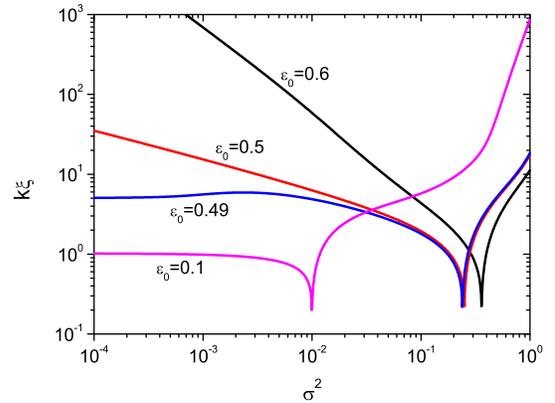}
 \caption{Normalized localization length $k\xi$ plotted versus disorder strength $\sigma^2$ in a log-log plot, when $kl_c=0.5$, $\theta=30^\circ$ and $\epsilon_0=0.1$, 0.49, 0.5, 0.6.}
 \label{fig:6}
\end{figure}

\begin{figure}
\centering\includegraphics[width=8.5cm]{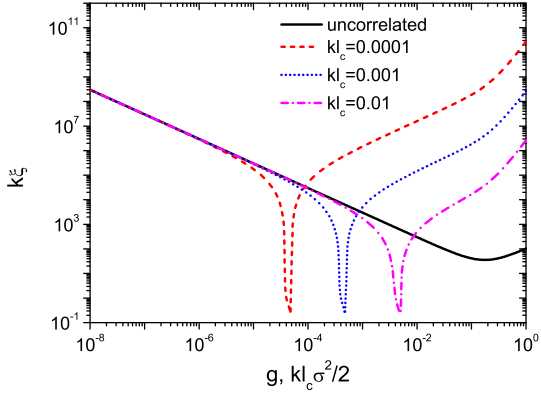}
\caption{Normalized localization length $k\xi$ plotted versus disorder strength $g$ (uncorrelated case)
or $kl_c\sigma^2/2$ (short-range correlated case) when $U_0=0$ and $\theta=30^\circ$.
The solid curve represents the uncorrelated case. The dashed and dotted curves represent the short-range correlated case for various values of
$kl_c$.}
\label{fig:7}
\end{figure}

We now consider the dependence of the localization length on the strength of disorder $\sigma^2$ in the short-range correlated model.
In Fig.~\ref{fig:5}, we plot the normalized localization length $k\xi$ versus $\sigma^2$ in a log-log plot,
when $kl_c=0.001$, $U_0=0$ and $\theta=30^\circ$. A very wide range of $\sigma^2$ from $10^{-4}$ to $10^6$ is considered.
First, we notice that the localization length has an extremely sharp dip at $\sigma^2=1$. We have carefully checked that the localization length
actually {\it vanishes} at this value, implying an extreme localization.
This unique phenomenon in pseudospin-1 systems arises due to the flat band located at $E=U$
and is directly related to the {\it singularity} of the wave equation, Eq.~(\ref{eq:we1}),
at $E=U$, or equivalently, to that of the invariant imbedding equation, Eq.~(\ref{eq:imbd1}), at $\epsilon=0$.
For our dichotomous random potential, the singularity condition corresponds to $E=U_0\pm S$, which is equivalent to $\epsilon_0=\pm\sigma$.
When $U_0$ is zero, this condition becomes $\sigma^2=1$.

We notice that there are three distinct scaling regions. In the weak disorder region where $\sigma^2\ll 1$, the localization length
decreases as $\xi\propto\sigma^{-2}$ as $\sigma$ increases from zero. This scaling behavior is independent of the size of the correlation length
as can be seen from Eqs.~(\ref{eq:llap1a}) and (\ref{eq:llap1b})
and agrees very well with the analytical formula, Eq.~(\ref{eq:bwkb}).
In the strong disorder region where $\sigma^2\gg 1$, we notice that $\xi$ increases monotonically as the disorder strength increases,
which implies that localization is destroyed by infinitely strong disorder, similarly to the pseudospin-1/2 case \cite{kkd}.
There are two different scaling behaviors
given by Eqs.~(\ref{eq:llap3a}) and (\ref{eq:llap3b}).
The crossover between them occurs at the value of $\sigma^2$ determined by Eq.~(\ref{eq:llap4}),
namely, $\sigma^2=\epsilon_0/(2kl_c)=500$ for the parameter values used here.
As we have discussed in Sec.~\ref{sec:ds}, in the region where $\xi\propto \sigma^2$, the effective wavelength is sufficiently larger than
the correlation length and the correlation effect is negligible. In contrast, in the asymptotically strong disorder region where
$\xi\propto \sigma^6$, the effective wavelength is much smaller than
the correlation length and the correlation effect becomes highly relevant.
The occurrence of the $\sigma^6$ scaling behavior and the crossover between the two scaling regions have not been
obtained before.

Next, we consider the case where $U_0$ is nonzero.
We first point out that the three scaling regions where $\xi$ is proportional to $\sigma^{-2}$, $\sigma^2$ or $\sigma^6$
also occur in this case, except in the total reflection regime where $\vert\epsilon_0\vert<\vert\sin\theta\vert$.
In Fig.~\ref{fig:6}, we plot the normalized localization length $k\xi$ versus $\sigma^2$ in a log-log plot,
when $kl_c=0.5$, $\theta=30^\circ$ and $\epsilon_0=0.1$, 0.49, 0.5 and 0.6. We verify that the sharp dips of the curves
occur precisely at the $\sigma^2$ values equal to ${\epsilon_0}^2$ as is expected. For the incident angle $\theta=30^\circ$,
the critical value of $\vert\epsilon_0\vert$ is equal to 0.5. We find that the localization length shows a nontrivial scaling behavior
of the form $\xi\propto \sigma^{-2/3}$ in the weak disorder region at $\epsilon_0=0.5$, whereas it scales as $\xi\propto \sigma^{-2}$ if $\epsilon_0>0.5$.
Contrasting scaling behaviors of this kind are observed in all systems showing the disorder-enhanced tunneling phenomenon \cite{kkd,kim7}.
We have performed a similar calculation for the $\delta$-correlated case and found that at the critical value, $\xi$ scales as $\xi\propto g^{-1/3}$,
which is equivalent to the result in the short-range correlated case. Precisely the same scaling behavior was previously obtained in the
pseudospin-1/2 case \cite{kkd}.

In contrast to the $\delta$-correlated case, the disorder-enhanced tunneling phenomenon in the present case does not occur in the whole total reflection regime, but is limited to
the parameter region where both Eq.~(\ref{eq:ie1}) and the condition $\vert\epsilon_0\vert<\vert\sin\theta\vert$ are satisfied.
For the parameter values used here, this gives the bound $0.455<\epsilon_0<0.5$.
For $\epsilon_0=0.49$, we find that $\xi$ initially increases to a maximum and then decreases as $\sigma$ increases, while
for $\epsilon_0=0.1$, $\xi$ decreases monotonically until the dip at $\sigma^2={\epsilon_0}^2$.

Finally, in Fig.~\ref{fig:7}, we compare the scaling behaviors in the weak disorder limit between the $\delta$-correlated case
and the short-range correlated case, when $U_0=0$ and $\theta=30^\circ$.
We find that the agreement between the two is perfect in the weak disorder limit, if we identify $g=kl_c\sigma^2/2$.
The sharp dips in the short-range correlated case occur precisely at $kl_c\sigma^2/2=kl_c/2$.
We observe that the result from the $\delta$-correlated model does not
show a sharp dip because our method in this case does not capture the singularity effect due to the flat band.

\subsection{Energy and wavelength dependence}

\begin{figure}
\centering\includegraphics[width=8.5cm]{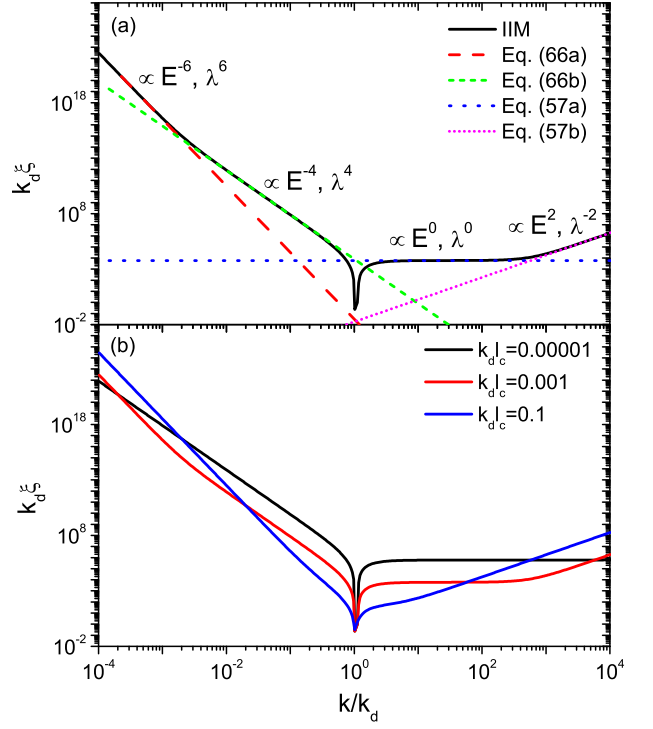}
 \caption{Energy dependence of the localization length in the short-range correlated case when the average potential $U_0$ is zero.
 $\xi$ is normalized by the wave number associated with disorder, $k_d$ [see Eq.~(\ref{eq:disk01})], and $k/k_d$
  ($=E/S$) is the normalized energy variable.
  In (a), the parameters used are $\theta=30^\circ$ and $k_dl_c=0.001$. The IIM result is compared with the analytical formulas, Eqs.~(\ref{eq:ed3a}), (\ref{eq:ed3b}), (\ref{eq:ed2a}) and (\ref{eq:ed2b}). (b) Normalized localization length $k_d\xi$
  plotted versus $k/k_d$ when $\theta=30^\circ$ and $k_dl_c=0.00001$, 0.001, 0.1.}
 \label{fig:8}
 \end{figure}

\begin{figure}
\centering\includegraphics[width=8.5cm]{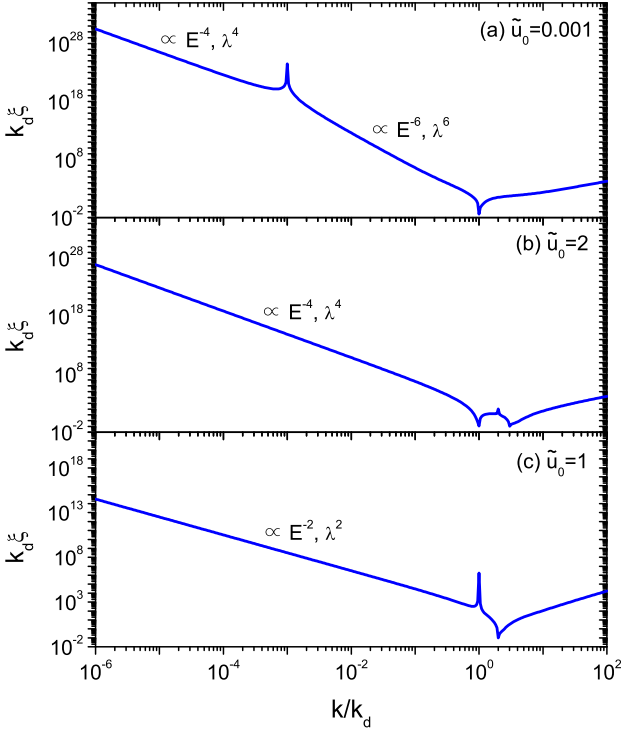}
 \caption{Energy dependence of the localization length in the short-range correlated case when the average potential $U_0$ is nonzero.
 The normalized localization length $k_d\xi$ is plotted versus $k/k_d$ when $\theta=30^\circ$, $k_dl_c=0.1$ and (a) $\tilde{u}_0=U_0/S=0.001$, (b) $\tilde{u}_0=2$, (c) $\tilde{u}_0=1$.}
 \label{fig:9}
 \end{figure}

\begin{figure}
\centering\includegraphics[width=8.5cm]{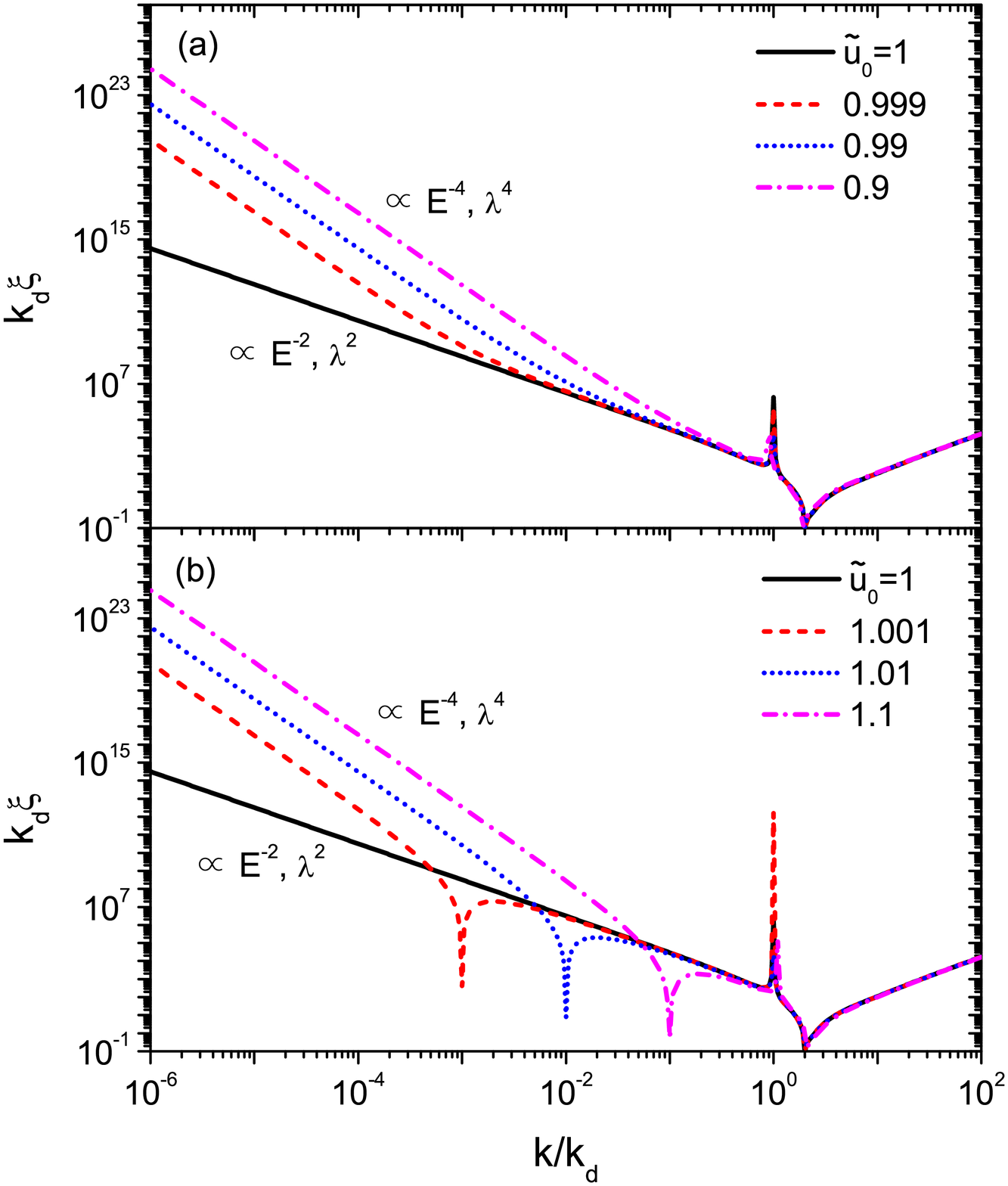}
 \caption{Energy dependence of the localization length in the short-range correlated case when the parameter $\tilde u_0$ ($=U_0/S$) is close to 1.
 The normalized localization length $k_d\xi$ is plotted versus $k/k_d$ when $\theta=30^\circ$, $k_dl_c=0.1$ and
 (a) $\tilde{u}_0\le 1$, (b) $\tilde{u}_0 \ge 1$.}
 \label{fig:10}
 \end{figure}

 \begin{figure}
\centering\includegraphics[width=8.5cm]{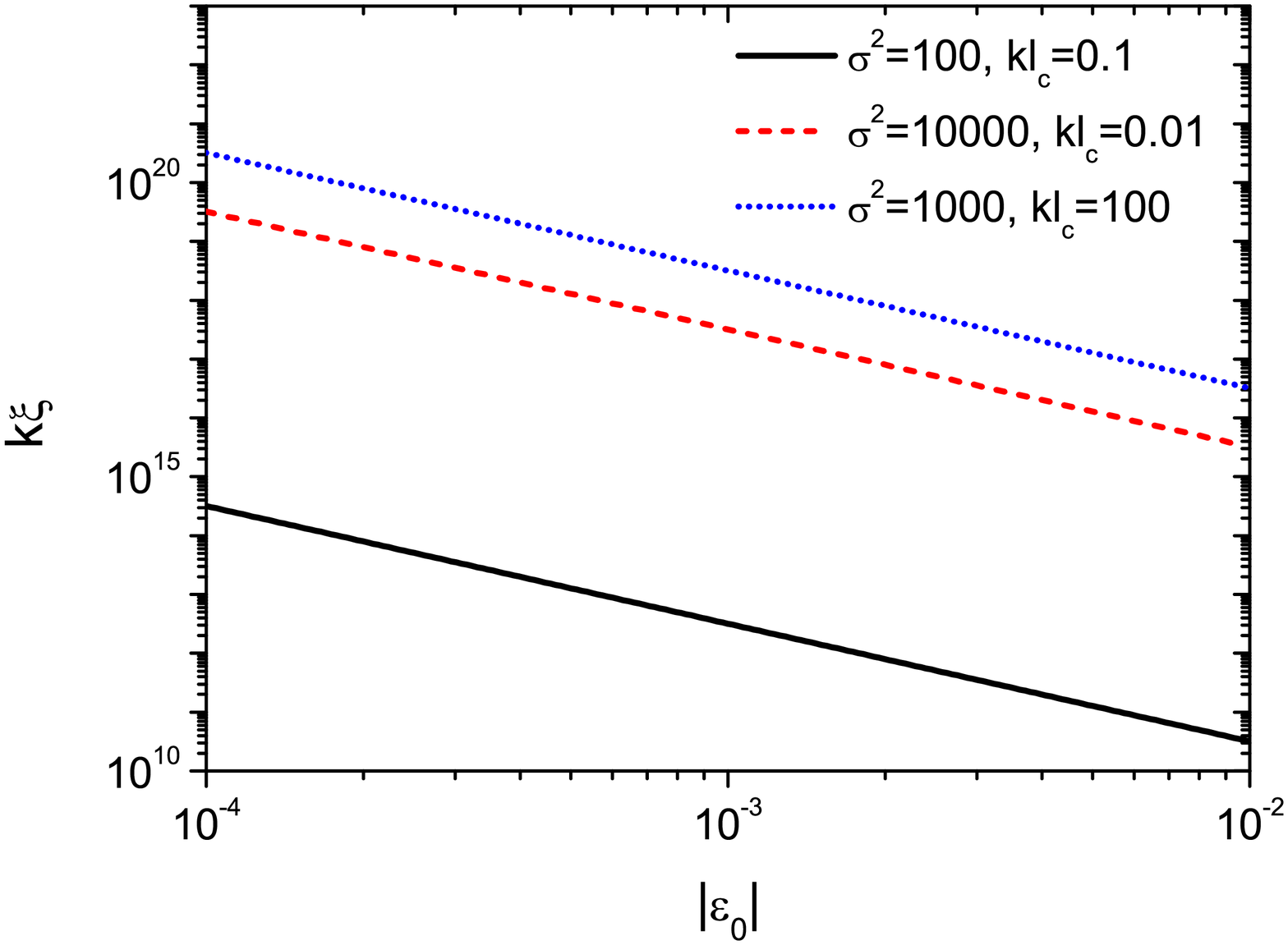}
\caption{Normalized localization length $k\xi$ in the short-range correlated case plotted versus $\vert\epsilon_0\vert$
 when $\theta=30^\circ$ in the strong disorder regime in a log-log plot. All curves show the divergent scaling behavior $k\xi\propto{\epsilon_0}^{-2}$.}
\label{fig:11}
\end{figure}

In Secs.~\ref{sec:dw2} and \ref{sec:ds}, we have discussed the energy and wavelength dependence of the localization length in
the weak and strong disorder regimes in the short-range correlated case.
We have found that when the average potential $U_0$ is zero, there should appear four different scaling regions depending on energy or wavelength.
In Fig.~\ref{fig:8}(a), we show the result of the IIM calculation when $k_dl_c=0.001$, $U_0=0$ and $\theta=30^\circ$.
The sharp dip occurs at the expected position $k/k_d=E/S=1$, which is obtained from the condition $E=U_0\pm S=\pm S$.
We see clearly that there are four scaling regions, where $\xi$ is proportional to $E^{-6}$, $E^{-4}$, $E^{0}$ or $E^{2}$, or equivalently,
to $\lambda^{6}$, $\lambda^{4}$, $\lambda^{0}$ or $\lambda^{-2}$. The IIM result is compared with
the analytical formulas, Eqs.~(\ref{eq:ed3a}), (\ref{eq:ed3b}), (\ref{eq:ed2a}) and (\ref{eq:ed2b}) and the agreements are quite good.

The crossover between different scaling behaviors is predicted to occur at $k/k_d=2k_dl_c=0.002$ in the small energy region
and at $k/k_d=1/(2k_dl_c\cos\theta)\approx 577.35$ in the large energy region and the curve shows a good agreement
with the predictions. As we have shown in Secs.~\ref{sec:dw2} and \ref{sec:ds}, both crossovers occur when the relative magnitudes
of the disorder correlation length and the effective wavelength change in the weak and strong disorder regimes respectively.
In the $E^{-4}$ and $E^{0}$ scaling regions, the effective wavelength is much larger than the correlation length
and the correlation effect is negligible, while in the $E^{-6}$ and $E^{2}$ scaling regions, the effective wavelength is much smaller than the correlation length and the correlation effect is important.

In Fig.~\ref{fig:8}(b), we compare the curves obtained for different values of $k_dl_c$. When $k_dl_c$ is 0.00001, the crossovers should
occur at $k/k_d=0.00002$ and $k/k_d\approx 57735$. Since these values are outside of the range shown here, we should observe only
the $E^{-4}$ and $E^{0}$ scaling behaviors, as can be verified from the figure. On the other hand, when $k_dl_c$ is 0.1,
the crossovers
occur at $k/k_d=0.2$ and $k/k_d\approx 5.77$.

In Sec.~\ref{sec:ds}, we have proved that when the average potential $U_0$ is nonzero, the scaling behavior
in the asymptotically small energy or long wavelength limit should be $\xi\propto E^{-4}$, $\lambda^4$ instead of $\xi\propto E^{-6}$, $\lambda^6$.
In Fig.~\ref{fig:9}, we plot the normalized localization length
$k_d\xi$ versus $k/k_d$ when $\theta=30^\circ$, $k_dl_c=0.1$ and the parameter $\tilde{u}_0$ ($=U_0/S$) takes the values 0.001, 2 and 1. We expect that sharp dips should occur at $k/k_d=\tilde u_0\pm 1$. This condition gives
$k/k_d=1.001$ in Fig.~\ref{fig:9}(a), $k/k_d=1, 3$ in Fig.~\ref{fig:9}(b) and $k/k_d=2$ in Fig.~\ref{fig:9}(c), which are confirmed in the figures.
In addition, we observe that sharp delocalization peaks appear when the condition $E=U_0$ is satisfied.
This condition is equivalent to $k/k_d=\tilde u_0$,
which is also confirmed in the figures. We have checked carefully that the localization length {\it diverges} precisely at $k/k_d=\tilde u_0$.

Except when $\tilde{u}_0$ is either 0 or 1, we confirm that the scaling behavior in the asymptotically small energy
limit is indeed given by $\xi\propto E^{-4}$. However, if $\tilde{u}_0$ is very close to zero as in Fig.~\ref{fig:9}(a),
$\xi$ follows the scaling
behavior $\xi\propto E^{-6}$ first and then crosses over to $\xi\propto E^{-4}$ as the energy decreases to zero.
The $\xi\propto E^{-2}$ scaling behavior shown in Fig.~\ref{fig:9}(c) for $\tilde u_0=1$ is peculiar and is different from the other cases.
Our perturbation theory in the weak and strong disorder regimes given in Secs.~\ref{sec:dw2} and \ref{sec:ds} cannot be applied to this case,
since the perturbation parameter $s$
is given by $s=\sigma^2/(1-\sigma)^2$, which approaches to 1 in the small energy limit where $\sigma$ diverges.
Therefore $s$ is neither large nor small and the perturbation expansion does not work.
Since $U_0$ is equal to $S$ when $\tilde u_0=1$, our dichotomous random potential fluctuates randomly between 0 and $2S$.
In Ref.~\cite{fang2}, it was reported that in a random superlattice structure where layers with zero potential and those
with a random potential were alternated periodically, the localization length scaled as $\xi\propto \lambda^2$ in the long wavelength limit.
Though this case and our case with $\tilde u_0=1$ are similar and show the same scaling behavior,
there is a subtle difference between them. In our case, the potential fluctuates randomly
between 0 and a constant value $2S$, while in Ref.~\cite{fang2}, it alternates periodically between 0 and a random value.
The common feature is the occurrence of the regions where the potential is identically equal to zero.

We elucidate the scaling behavior in the $\tilde u_0=1$ case further by calculating the localization length for $\tilde u_0$
slightly different from 1.
In Fig.~\ref{fig:10}, we plot the normalized localization length $k_d\xi$ versus $k/k_d$ when $\theta=30^\circ$, $k_dl_c=0.1$ and
$\tilde{u}_0$ is very close to 1. As is expected, sharp dips occur at $\tilde u_0\pm 1$ and sharp peaks occur at $\tilde u_0$. We find that as the energy decreases to zero,
the scaling follows $\xi\propto E^{-2}$ first and then crosses over to $\xi\propto E^{-4}$.
Therefore the $E^{-2}$ scaling behavior strictly occurs only when $\tilde u_0=1$.

Finally, in Fig.~\ref{fig:11}, we show how the localization length diverges when $E$ approaches $U_0$, or equivalently, when $\epsilon_0$
approaches zero for several parameter values. We find that $\xi$ always diverges as $\xi\propto{\epsilon_0}^{-2}$.
The sharp delocalization peaks appearing in Figs.~\ref{fig:9} and \ref{fig:10}
obey this behavior. A similar result was reported in Ref.~\cite{fang2}.

\section{Conclusion}
\label{sec:con}

In this paper, we have studied the Anderson localization of 2D massless pseudospin-1 Dirac particles
in a random 1D scalar potential theoretically.
We have explored the effect of disorder correlations by solving the Dirac equation with
short-range correlated dichotomous random potential for all strengths of disorder.
Using the invariant imbedding method, we have calculated the localization length in a numerically exact manner and analyzed its dependencies
on incident angle, disorder correlation length, disorder strength, energy, wavelength and average potential extensively
over a wide range of parameter values.
We have also derived concise analytical expressions for the localization length, which are extremely accurate in the weak and strong
disorder regimes.
Using the effective wave impedance derived from the pseudospin-1 Dirac equation, we have obtained several delocalization
conditions for which the localization length diverges.
We have also obtained a condition for which the localization length vanishes.
For all cases studied in this paper, we have found that the localization length depends non-monotonically on the correlation length
and diverges as $\theta^{-4}$ at normal incidence.
As the disorder strength and the energy (or wavelength) vary from zero to infinity, we have found that
there appear several different types of scaling behaviors with different values of the scaling exponent.
We have explained the crossover between different scaling behaviors in terms of the relative magnitude of the correlation length
and the effective wavelength.

We hope our results will stimulate future experiments on localization in pseudospin-1 systems.
Our approach can be easily adapted to the case where both the random scalar and vector potentials are present.
It is also straightforward to apply our method to other pseudospin-$N$ systems such as pseudospin-3/2 and pseudospin-2 systems.
These directions of research will be pursued in the future.

\acknowledgments
This research was supported by the Basic Science Research Program through a National Research Foundation of Korea Grant (NRF-2019R1F1A1059024) funded by the Ministry of Education.

\end{document}